\documentclass[10pt,journal,compsoc]{IEEEtran}
\usepackage{verbatim}
\usepackage{acronym}
\usepackage{makecell} 
\usepackage{comment}
\usepackage{cite}
\usepackage[pdftex]{graphicx}
\usepackage{hyperref}
\hypersetup{
     colorlinks = true,
     linkcolor = blue,
     anchorcolor = blue,
     citecolor = blue,
     filecolor = blue,
     urlcolor = blue
}
\usepackage{pifont}
\usepackage{lscape}
\usepackage{longtable}
\newcommand{\cmark}{\ding{51}}
\newcommand{\xmark}{\ding{55}}

\usepackage{background}
\backgroundsetup{
  position=current page.east,
  angle=-90,
  nodeanchor=east,
  vshift=-5mm,
  opacity=1,
  scale=1,
  contents= Accepted for Publication in IEEE Access DOI: \href{10.1109/ACCESS.2020.3037715}{10.1109/ACCESS.2020.3037715}
}

\begin{document}
\bstctlcite{IEEEexample:BSTcontrol}

\acrodef{HAR}{Human Activity Recognition}
\acrodef{CNN}{Convolutional Neural Network}
\acrodef{LSTM}{Long Short-Term Memory}
\acrodef{RNN}{Recurrent Neural Network}
\acrodef{ELM}{Extreme Learning Machine}
\acrodef{DBN}{Dynamic Bayesian Network}
\acrodef{INN}{Inception Neural Networks}
\acrodef{SVM}{Support Vector Machine}
\acrodef{kNN}{k-Nearest Neighbor}
\acrodef{DT}{Decision Tree}
\acrodef{RF}{Random Forest}
\acrodef{NN}{Neural Networks}
\acrodef{NB}{Naïve Bayes}
\acrodef{HMM}{Hidden Markov Models}
\acrodef{PCA}{Principal Component Analysis}
\acrodef{LDA}{Linear Discriminant Analysis}
\acrodef{QDA}{Quadratic Discriminant Analysis}
\acrodef{ML}{Machine Learning}
\acrodef{CML}{Classic Machine Learning}
\acrodef{DL}{Deep Learning}
\acrodef{AI}{Artificial Intelligence}
\acrodef{RFID}{Radio Frequency Identification}
\acrodef{GPS}{Global Positioning System}
\acrodef{IoT}{Internet of Things}
\acrodef{WLAN}{Wireless Local Area Network}
\acrodef{BLE}{Bluetooth Low Energy}
\acrodef{DLO}{Daily Life Objects}
\acrodef{IDC}{International Data Corporation}
\acrodef{ADL}{Activities of Daily Life}
\acrodef{EMG}{Electromyography}
\acrodef{ECG}{Electrocardiogram}
\acrodef{RSSI}{Received Signal Strength Indicator}
\acrodef{EEG}{Electroencephalogram}
\acrodef{PRISMA}{Preferred Reporting Items for Systematic reviews and Meta-Analyses}
\acrodef{EOG}{Electrooculogram}
\acrodef{AAL}{Ambient Assisted Living}
\acrodef{HCR}{Human Context Recognition}
\acrodef{MEMS}{Microelectromechanical systems}
\acrodef{AR}{Activity Recognition}
\acrodef{ICT}{Information and Communication Technology}
\title{Human Activity Recognition using Inertial, Physiological and Environmental Sensors: A Comprehensive Survey}

\author{Florenc~Demrozi,~\IEEEmembership{Member,~IEEE,}
        Graziano~Pravadelli,~\IEEEmembership{Senior Member,~IEEE,}
        Azra~Bihorac,
        and,~Parisa~Rashidi,~\IEEEmembership{Senior Member,~IEEE,}
\thanks{F.Demrozi and G.Pravadelli~are with the Department
of Computer Science, University of Verona, Italy e-mail: name.surname@univr.it}
\thanks{P. Rashidi is with the Department of Biomedical Engineering, University of Florida, Gainesville, FL, USA 
e-mail: parisa.rashidi@ufl.edu}
\thanks{A. Bihorac is with the Division of Nephrology, Hypertension, \& Renal Transplantation, College of Medicine, University of Florida, Gainesville, FL, USA e-mail: abihorac@ufl.edu}
}



\IEEEtitleabstractindextext{%
\begin{abstract}
In the last decade, Human Activity Recognition (HAR) has become a vibrant research area, especially due to the spread of electronic devices such as smartphones, smartwatches and video cameras present in our daily lives. In addition, the advance of deep learning and other machine learning algorithms has allowed researchers to use HAR in various domains including sports, health and well-being applications. For example, HAR is considered as one of the most promising assistive technology tools to support elderly’s daily life by monitoring their cognitive and physical function through daily activities. This survey focuses on critical role of machine learning in developing HAR applications based on inertial sensors in conjunction with physiological and environmental sensors.
\end{abstract}

}
\maketitle
\IEEEdisplaynontitleabstractindextext
\IEEEpeerreviewmaketitle
\section{Introduction}\label{sec:intro}
HUMAN ACTIVITY RECOGNITION (HAR) has become a popular topic in the last decade due to its importance in many areas, including health care, interactive gaming, sports, and monitoring systems for general purposes~\cite{antunes2018survey}.
Besides, nowadays, the aging population is becoming one of the world’s primary concerns. It was estimated that the population aged over 65 would increase from 461 million to 2 billion by 2050. This substantial increase will have significant social and health care consequences. To monitor physical, functional, and cognitive health of older adults in their home, HAR is emerging as a powerful tool~\cite{wang2019survey}

The goal of HAR is to recognize human activities in controlled and uncontrolled settings. Despite myriad applications, HAR algorithms still face many challenges, including 1) complexity and variety of daily activities, 2) intra-subject and inter-subject variability for the same activity, 3) the trade-off between performance and privacy, 4) computational efficiency in embedded and portable devices, and 5) difficulty of data annotation~\cite{lara2012survey}. 
Data for training and testing HAR algorithms is typically obtained from two main sources, 1) ambient sensors, and 2) embedded sensors. 
Ambient sensors can be environmental sensors such as temperature sensors or video cameras positioned in specific points in the environment~\cite{liu2016action,zeng2014convolutional}. 
Embedded sensors are integrated into personal devices such as smartphones and smartwatches, or are integrated into clothes or other specific medical equipment~\cite{stisen2015smart,kanjo2019deep,neverova2016learning,liu2015sensor}. 
Cameras have been widely used in the HAR applications, however collecting video data presents many issues regarding privacy and computational requirements~\cite{donahue2015long}.
While video cameras produce rich contextual information, privacy issues limitations have led many researchers to work with other ambient and embedded sensors, including depth images as a privacy-preserving alternative.

\begin{figure*}[th!]
\minipage{0.45\textwidth}
\includegraphics[width=\linewidth,page={1}]{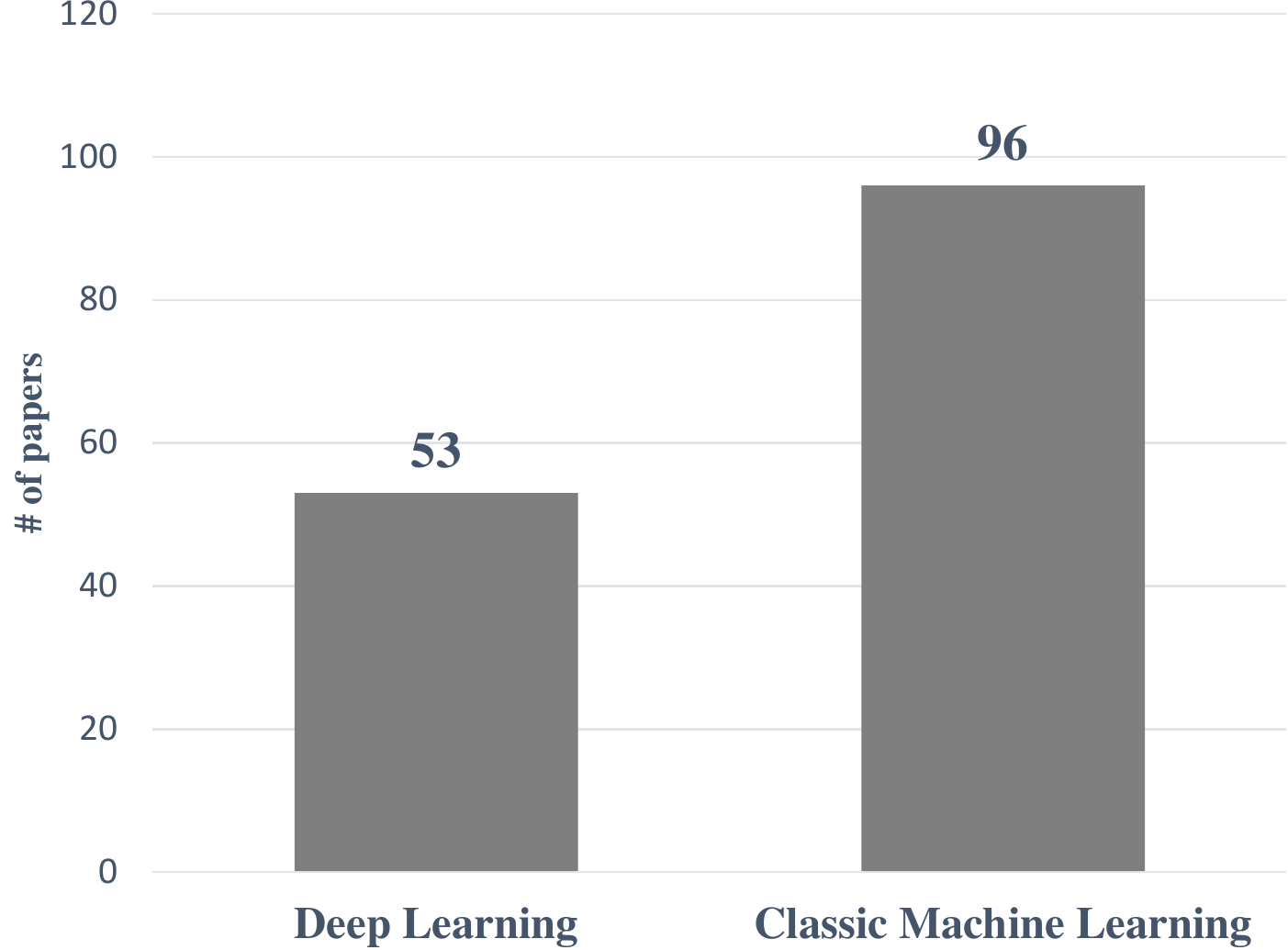} \\ \centering(a)
\endminipage\hfill
\minipage{0.45\textwidth}
\includegraphics[width=\linewidth,page={2}]{imm/p_figures.pdf} \\ \centering(b)
\endminipage\hfill 
\caption{(a) Distribution of published papers in HAR research area, for DL vs. CML implementations.  (b) The average recognition accuracy of published papers, for DL vs. CML implementations.}\label{fig:one}
\end{figure*}

\begin{figure*}[!bht]
\minipage{0.45\textwidth}
\includegraphics[width=\linewidth,page={3}]{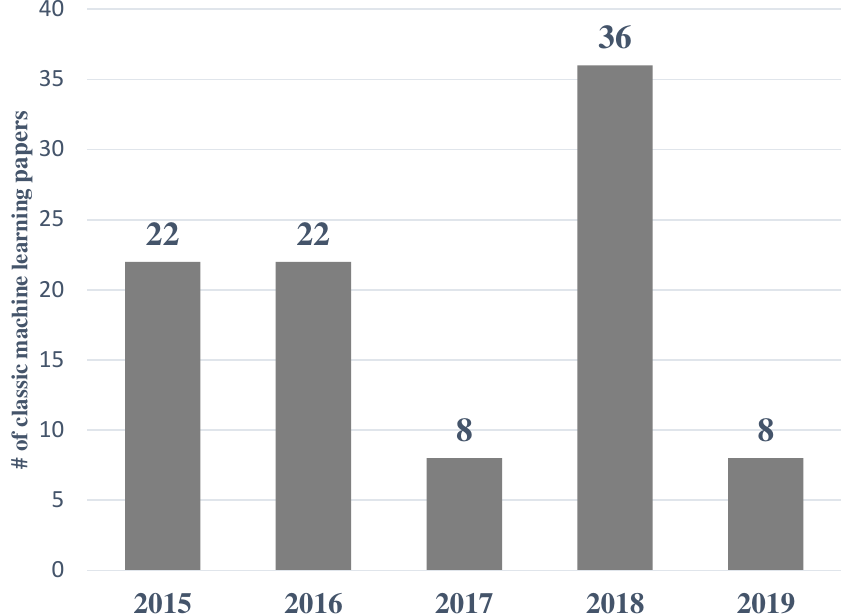} \\ \centering(a)
\endminipage\hfill
\minipage{0.45\textwidth}
\includegraphics[width=\linewidth,page={4}]{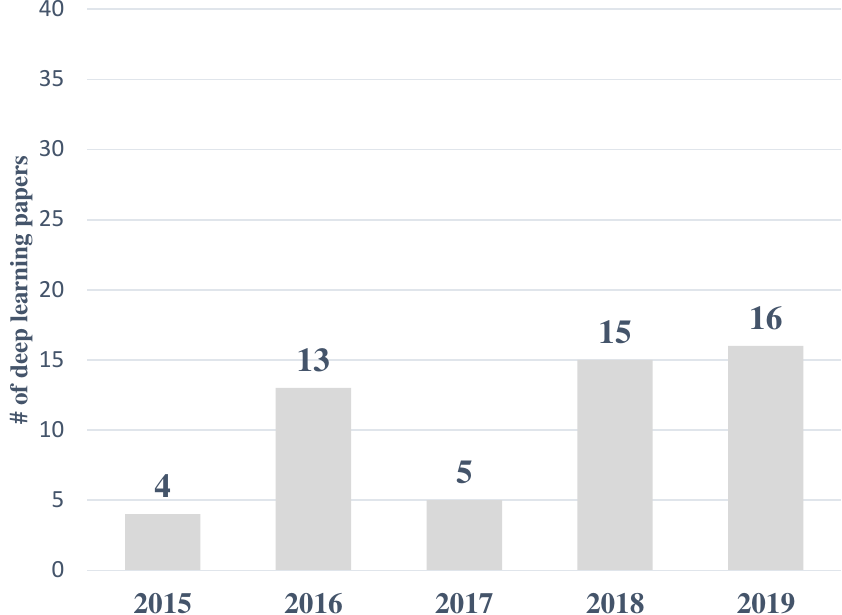} \\ \centering(b)
\endminipage\hfill 
\caption{Distribution of published papers per year in HAR research based on (a) CML and (b) DL.}\label{fig:three}
\end{figure*}

In terms of algorithmic implementation, HAR research has seen an explosion in Deep Learning (DL) methods, resulting in an increase in recognition accuracy~\cite{zeng2014convolutional,kanjo2019deep}. 
While DL methods produce high accuracy results on large activity datasets, in many HAR  applications Classic Machine Learning (CML) models might be better suited due to the small size of the dataset, lower dimensionality of the input data, and availability of expert knowledge in formulating the problem~\cite{lecun2015deep}. 
The increasing interest in HAR can be associated with growing use of sensors and wearable devices in all aspects of daily life, especially with respect to health and well-being applications. This increasing interest in HAR is evident from the number of papers published in the past five years, from 2015 to 2019.
As Figure~\ref{fig:one}.(a) shows, among a total of 149 selected published papers on HAR, 53 were based on DL models, and 96 were based on CML models. During the same time period were published 46 surveys and 20 articles proposing not ML-based methodologies (e.g., threshold models). 
Figure~\ref{fig:one}.(b) shows the average activity recognition accuracy, among the 53 DL-based papers and the 96 CML-based papers, that as visible (93\% DL-based and 92.2\% CML-based) present almost the same recognition quality.
In addition, Figure~\ref{fig:three} shows the distribution of the published HAR papers over the past five years in terms of (a) CML and (b) DL models. It shows that the number of CML-based HAR models was, except 2019, greater than the number of DL-based HAR models. 
In this paper, we will review both DL-based and CML-based methodologies. We will limit our review to non-image-based sensors, to limit the scope. Interested readers are encouraged to read references on vision-based HAR~\cite{donahue2015long,wang2015action,liu2016spatio,voulodimos2018deep}.

\begin{figure*}[!ht]
\centering
\includegraphics[width=0.8\textwidth,page={7}]{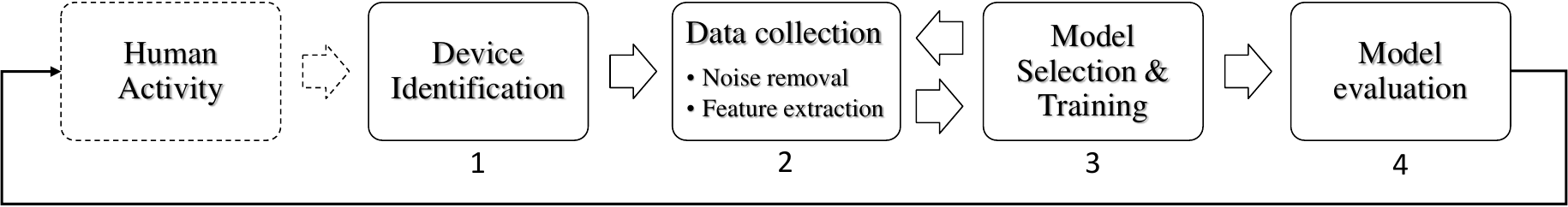}
\caption{Standard workflow for implementing HAR based application.}\label{fig:two}
\end{figure*}

Figure~\ref{fig:two} presents the standard workflow in designing HAR-based methodologies. When developing HAR-based application, the first step is to determine the type of sensor and device that is used to collect data (device identification). The second step is to determine the details of the data collection process, including the annotation process and possibly any necessary preprocessing (data collection). The third step includes identifying the appropriate machine learning model and training the model, typically a supervised machine learning model on annotated data (model selection and training). However, as shown in Figure~\ref{fig:two} (indicated by the backwards arrow), the selected model can also influence the preprocessing data step. In the final step, the model is evaluated in terms of the activity recognition metrics such as accuracy, precision, recall, and other metrics (model evaluation). 
In this work, we use accuracy as a comparison metric between the various articles due to the fact that it is the only common metric. Not all articles present the results obtained in terms of precision, recall, sensitivity, F1-Score, Area Under the Curve (AUC) or Receiver Operating Characteristics (ROC) curve, despite being more representative metrics, especially with unbalanced data. Using this workflow as a reference, this paper provides an overview of the state-of-the-art in HAR by examining each phase of the process. Finally, we are particularly interested in accelerometer sensors because they have shown excellent results in HAR applications and because their use in conjunction with other sensors is rising rapidly. The proliferation of accelerometer sensors is strongly related to their ability to measure directly the movement of the human body. In addition, using accelerometer sensors is affordable, and the sensors can be integrated into most wearable electronic objects people own.\\

\noindent The rest of the paper is organized as follows: 
Section~\ref{sec:surv} provides a brief overview of the existing surveys on HAR from 2015 to 2019,  
Section~\ref{sec:sel} describes the article selection criteria, Section~\ref{sec:back} will provide background material on CML, DL, and existing sensors/wearable devices. 
Section~\ref{sec:humact}  will introduce the definition of human activity, followed by  categorization of the published works in terms of sensor and device (Section~\ref{sec:device}). 
Section~\ref{sec:data} will present available datasets for HAR research activity. Section~\ref{sec:eval} will review published papers based on the model and evaluation metrics. Section~\ref{sec:disc} will discuss the limitations and challenges of existing HAR research, followed by a discussion on future research direction in Section~\ref{sec:future}. Finally, Section~\ref{sec:conc} reports some concluding remarks.
\section{Existing Surveys}\label{sec:surv}
Since HAR is emerging as an important research topic, many surveys have been published in the past few years. Among the initial 293 published papers that we identified, 46 were survey papers published since 2015. The existing survey papers can be categorized based on the data sources and the activity recognition algorithm. The most widely used data sources are a) inertial, physiological and environmental devices, and b) video recording devices. In terms of the HAR algorithm, most algorithms are based on CML models and more recently DL algorithms. Among such 46 survey papers, we excluded 23 papers which were exclusively video-based HAR papers. 
Our survey paper provides unique contribution to the review of literature by providing, a broad vision of the evolution of HAR research in the past 5 years. Unlike existing surveys, we do not solely focus on the algorithmic details, rather we will also describe the data sources (aka sensors and devices) are used in this context. We are particularly interested in accelerometer sensors because they have shown excellent results in HAR applications and because their use in conjunction with other sensors such as physiological sensors or environmental sensors is rising rapidly. The proliferation of accelerometer sensors is strongly related to their ability to directly measure the movement of the human body. In addition, using accelerometer sensors is affordable, and the sensors can be integrated into most wearable devices. 
Recently, Wang. J and colleagues~\cite{wang2019deep} (2019) survey existing literature based on three aspects: sensor modality, DL models, and application scenarios, presenting detailed information of the reviewed works. Wang. Y and colleagues~\cite{wang2019survey} (2019) present the state-of-the-art sensor modalities in HAR mainly focusing on the techniques associated with each step of HAR in terms of sensors, data preprocessing, feature learning, classification, activities, including both conventional and DL methods. Besides, they present the ambient sensor-based HAR, including camera-based, and systems combining wearable and ambient sensors. 
Sousa et al.~\cite{sousa2019human} (2019) provide a complete, state-of-the-art outline of the current HAR solutions in the context of inertial sensors in smartphones, and, Elbasiony et al.~\cite{elbasiony2019survey} (2019) introduce a detailed survey on multiple HAR systems on portable inertial sensors (Accelerometer, Gyroscopes, and Magnetometer),  whose temporal signals are used for modeling and recognition of different activities.

Nweke et al.~\cite{nweke2019data} (2019) provide a detailed analysis of data/sensors fusion and multiple classification systems techniques for HAR with emphasis on mobile and wearable devices. 
Faust et al.~\cite{faust2018deep} (2018), studied 53 papers focused on physiological sensors used in healthcare applications such as  \ac{EMG}, \ac{ECG}, \ac{EOG}, and \ac{EEG}. Ramasamy~\cite{ramasamy2018recent} (2018) presented an overview of ML and data mining techniques used for Activity Recognition (AR), empathizing with the fundamental problems and challenges.
Finally, Morales et al.~\cite{morales2017physical} (2017) provide an overview of the state-of-the-art concerning: relevant signals, data capturing and preprocessing, calibrating on-body locations and orientation, selecting the right set of features, activity models and classifiers, and ways to evaluate the usability of a HAR system. Moreover, it covers the detection of repetitive activities, postures, falls, and inactivity.

\noindent Table~\ref{tab:surveys} summarizes 23 surveys on HAR methods sorted by chronological order from 2019 to 2015. It should be noted that all these surveys, including those not taken into consideration (video-based), had not reported their systematic review process (e.g., using Preferred Reporting Items for Systematic reviews and Meta-Analyses (PRISMA)). In Table~\ref{tab:surveys}, Column five, we report the start/end publication year of the reviewed papers and Column six their approximate number of reviewed articles. Most of these HAR reviews focus on data management methods and activity recognition models. To the best of our knowledge, no existing survey article is (1) presenting a comprehensive meta-review of the existing surveys, (2) providing a comprehensive overview of different sensors, (3) reporting and comparing performance metrics, and (4) reporting on dataset availability and popularity.

\begin{table*}[!ht]
\caption{Existing HAR Surveys}\label{tab:surveys}
\centering
\resizebox{0.95\textwidth}{!}{
\begin{tabular}{c c p{2cm} p{7cm} c c c}
\hline
\textbf{Reference}  & \makecell{\textbf{Publication}\\\textbf{Year}} & \makecell{\textbf{Main}\\ \textbf{Focus}}     & \makecell{\textbf{Used}\\\textbf{Keywords}}    & \makecell{\#\\\textbf{ Keywords}}              & \makecell{\textbf{Start/End}\\\textbf{Year}}   &
\makecell{\# \textbf{Reviewed}\\\textbf{Papers}}\\
\hline
\hline
\cite{wang2019deep}&2019&DL based HAR&DL, Activity Recognition, Pattern Recogniton, Pervasive Computing&4&2013/2019&77\\
\hline
\cite{wang2019survey}&2019&HAR in HealthCare&HAR, Wearables sensors, DL, Features, Healthcare&5&2005/2019&258\\
\hline
\cite{sousa2019human}&2019&Inertial Sensors in Smartphones based HAR&HAR, activity recognition, smartphones, mobile phones, inertial sensors, accelerometer, gyroscope, \ac{ML}, classification algorithms, \ac{DL}&10&2006/2019&149\\
\hline
\cite{elbasiony2019survey}&2019&Temporal Signals based HAR&HAR, ML, Inertial measurement unit, Accelerometer, Gyroscope&6&2001/2019&48\\
\hline
\cite{nweke2019data}&2019&HAR on multi data system&Activity detection, Data fusion , DL, Health monitoring, Multiple classifier systems, Multimodal sensors&6&2005/2019&309\\
\hline 
\cite{nweke2018deep}&2018&Smartphones based HAR&DL, Mobile and wearable sensors,  HAR, Feature representation&4&2005/2018&275\\
\hline
\cite{faust2018deep}&2018&DL for health on physiological signals&DL, Physiological signals, Electrocardiogram, Electroencephalogram, Electromyogram, Electrooculogram&6&2008/2017&53\\
\hline
\cite{ramasamy2018recent}&2018&ML based HAR&active learning, activity recognition, data mining, DL, ML, transfer learning, wearable sensors&7&2007/2018/87\\
\hline
\cite{keng2018review}&2018&HAR for Ageing&Senior citizens, Activity recognition, Internet of Things, Intelligent sensors, Aging, Task analysis&6&2010/2018&43\\
\hline
\cite{miotto2018deep}&2017&DL for healthcare&DL, health care, biomedical informatics, translational bioinformatics, genomics, electronic health records&6&2012/2017&119\\
\hline
\cite{gamboa2017deep}&2017&Time Series based DL&Artificial Neural Networks, \ac{DL}, Time-Series&3&2007/2017&60\\
\hline
\cite{dhillon2017recent}&2017&DL based HAR&DL, Activity Recognition, Video, Motion&4&2010/2017&24\\
\hline
\cite{chen2017survey}&2017&Video and Inertials based HAR&HAR, Activity recognition, 3D action data, Depth sensor, Inertial sensor, Sensor fusion, Multimodal dataset&7&2010/2017&78\\
\hline
\cite{morales2017physical}&2017&HAR with Smartphones&Accelerometer, Gyroscope, \ac{AR}, Smartphone&4&2010/2017&64\\
\hline
\cite{vyas2017survey}&2017&Smartphones based HAR &\ac{AR}, Sensors, Smartphone, Activity of Daily Living, Aceelerometer, Survey, Processing&7&2005/2017&39\\
\hline
\cite{rault2017survey}&2017&Smartphones based HAR&State-of-the-art, Energy efficient wearable sensor networks, Human context recognition&3&2007/2017&88\\
\hline
\cite{patel2017human}&2017&HAR general overview&artificial intelligent, human body posture recognition, feature extraction, classification&4&2010/2017&13\\
\hline
\cite{kumari2017increasing}&2017&Wearable based HAR&Human activity monitoring, Human Computer Interface, Wearable sensors, Smart sensors, Multimodal interface, Biomedical, Shared control architecture&7&2005/2016&85\\
\hline
\cite{ravi2016deep}&2016&DL for healthcare&Bioinformatics, DL, health informatics, ML, medical imaging, public health, wearable devices&7&2010/2016&145\\
\hline
\cite{cornacchia2016survey}&2016&Wearable based HAR&Wearable, sensors, survey, activity detection, activity classification, monitoring&5&2005/2016&225\\
\hline
\cite{woznowski2016classification}&2016&Smartphone based HAR&\ac{AR}, Sensors, ADL&3&2001/2015&138\\
\hline
\cite{shoaib2015survey}&2015&Wearable based HAR&online \ac{AR}, real time, smartphones, mobile phone, mobile phone sensing, \ac{HAR} review, survey, accelerometer&8&2008/2014&74\\
\hline
\cite{ciuti2015mems}&2015&HAR in Health-Care& MEMS sensor technologies, human centered applications, research activity in Italy, healthcare, rehabilitation, physical activities, sport science, safety, environmental sensing&9&2011/2014&128\\
\hline
\hline
\end{tabular}}
\end{table*}
\section{Selection Criteria}\label{sec:sel}
\begin{figure}[!bh]
\centering
\includegraphics[width=0.4\textwidth,page={8}]{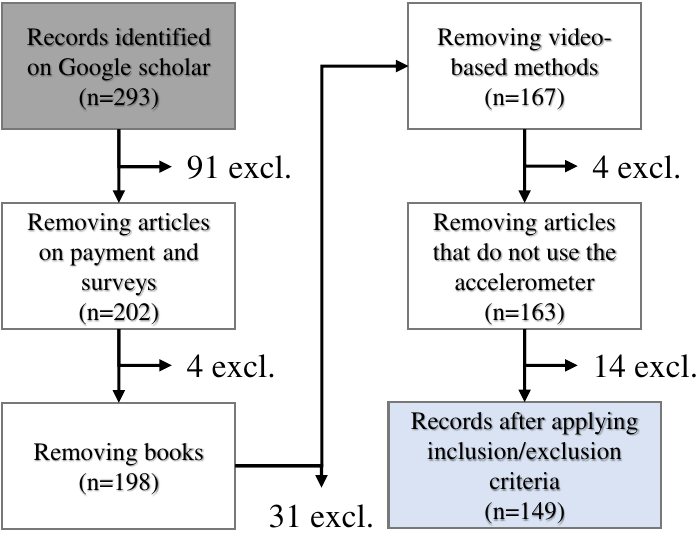}
\caption{PRISMA-based flowchart of the retrieval process.}\label{fig:four}
\end{figure}

\begin{table}[!ht]
\caption{Distribution of the selected published articles for year by including the following keywords: "\acf{HAR}, Sensor $<$Name$>$, Wearable sensors".}\label{tab:one}
\centering
\resizebox{0.45\textwidth}{!}{
\begin{tabular}{c|ccccc|c}
\hline
Year & 2015 & 2016 & 2017& 2018 & 2019 & Total \\ \hline
Total \# of Papers & 52 & 60 & 45 & 90 & 46 & 293 \\ 
\hline
\hline
\end{tabular}
}
\end{table}
We used Google Scholar to search for studies published between January 2015 to September 2019. All searches included the term “human activity recognition,” or “HAR” in combination with “deep learning”, “machine learning,” “wearable sensors,” and “$<$name$>$\footnote{e.g., accelerometer, gyroscope, magnetometer, barometer, light, \ac{GPS}} sensor”. 
All these published papers where found by using the combination of keywords mentioned above.
Our keywords produced a total of 249110 records, among which we selected 293 based on the quality of the publication venue. The chosen articles were selected from the following publishers: Institute of Electrical and Electronics Engineers (IEEE), Association for Computing Machinery (ACM), Elsevier, and Sensors. The average number of citations was 46, and the distribution of the papers for each year is shown in Table~\ref{tab:one}. Figure~\ref{fig:four} shows our retrieval process based on PRISMA template for systematic reviews~\cite{moher2009preferred}.
First, we excluded all surveys papers and not accessible papers ( e.g., requiring paid access) (91 excluded). Next, we excluded all books (4 excluded) and all vision-based papers (31 excluded). Finally, we excluded all the papers that do not use accelerometers (4 excluded), and all the papers performing activity recognition different from daily life human activities, such as swimming, riding horses, driving, publications prior to 2015, and papers using non-machine learning techniques such as simple thresholding (4 excluded). As a result, 149 were eligible, as Figures~\ref{fig:one} and~\ref{fig:four} show. 
\section{Background}\label{sec:back}
The main objective of HAR algorithms is to recognize human activity based on data gathered by wearable and environmental sensors~\cite{wang2019deep,demrozi2019towards}. 
The recognition of these activities is mainly based on CML and DL  algorithms.
Recently, the use of a wide variety of sensors, has generated interest in sensor fusion techniques. 
This section introduces basic ML and DL concepts, wearable/environmental sensors market evolution, and sensor fusion techniques.

\subsection{Machine Learning Overview}
Machine Learning (ML) is a branch of Artificial Intelligence (AI), for developing algorithms that can identify and infer patterns given a training dataset~\cite{bishop2006pattern}. Such algorithms fall into two major classes:
\begin{itemize}
    \item Supervised learning, 
    \item Unsupervised learning. 
\end{itemize}
The goal of supervised learning is to create a mathematical model based on the relationship between input and output data and to use the model for predicting future unseen data points. 
In unsupervised learning, the goal is to identify patterns in input data without any knowledge of the output~\cite{liu2016action}. 
Typically, one or more preprocessing steps will be also required, including feature extraction, vectorization/segmentation, normalization or standardization, and projection~\cite{domingos2012few}. \\
Some of the most common supervised CML algorithms are: Naïve Bayes (NB), k-Means Clustering, Support Vector Machine (SVM), Linear Regression, Logistic Regression, Random Forests (RF), Decision Trees (DT) and k-Nearest Neighbours (k-NN). 
DT's classify data instances by sorting them based on the features/data values. Each node represents a feature to be classified, and each branch represents a value that the node can assume. 
NB classifiers are probabilistic classifiers based on applying Bayes' theorem with strong independence assumptions between the features. 
SVMs are based on the notion of a margin-either side of a hyperplane that separates two data classes. Maximizing the margin, thereby creating the most significant possible distance between the separating hyperplane and the instances on either side, has been proven to reduce an upper bound on the expected generalization error.
Finally, K-NN is a CML algorithm that stores all available cases and classifies new cases based on a similarity measure (e.g., distance functions as Euclidean, Manhattan, Minkowski)\cite{bishop2006pattern}. 
Furthermore, since HAR imposes specific constraints, such as reduced latency, memory constraint, and computational constraints, these classifiers, except for SVM, are appropriate for low-resource environments given their low computational and memory requirements.

Among unsupervised and particularly clustering algorithms, the most well-known algorithms are k-Means, Hierarchical clustering, and Mixture models. 
K-Means clustering aims to partition groups of samples into k clusters based on a similarity measure (intra-group) and dissimilarity measure (inter-groups). 
Each sample belongs to the cluster with the nearest cluster centers or cluster centroid, serving as a cluster prototype. Hierarchical Clustering Analysis is a cluster analysis method that seeks to build a hierarchy of clusters where clusters are combined/split based on the measure of dissimilarity between sets.
A mixture model is a probabilistic model for representing subpopulations of observations within an overall population ~\cite{liu2016action}.
These techniques  are particularly suitable when working with datasets lacking labels or when the measure of similarity/dissimilarity between classes is a primary outcome~\cite{dobbins2018towards,vaughn2018activity,abdallah2015adaptive}.

\subsection{Deep Learning Overview}
\begin{figure}
\centering
\includegraphics[width=0.4\textwidth,page={21}]{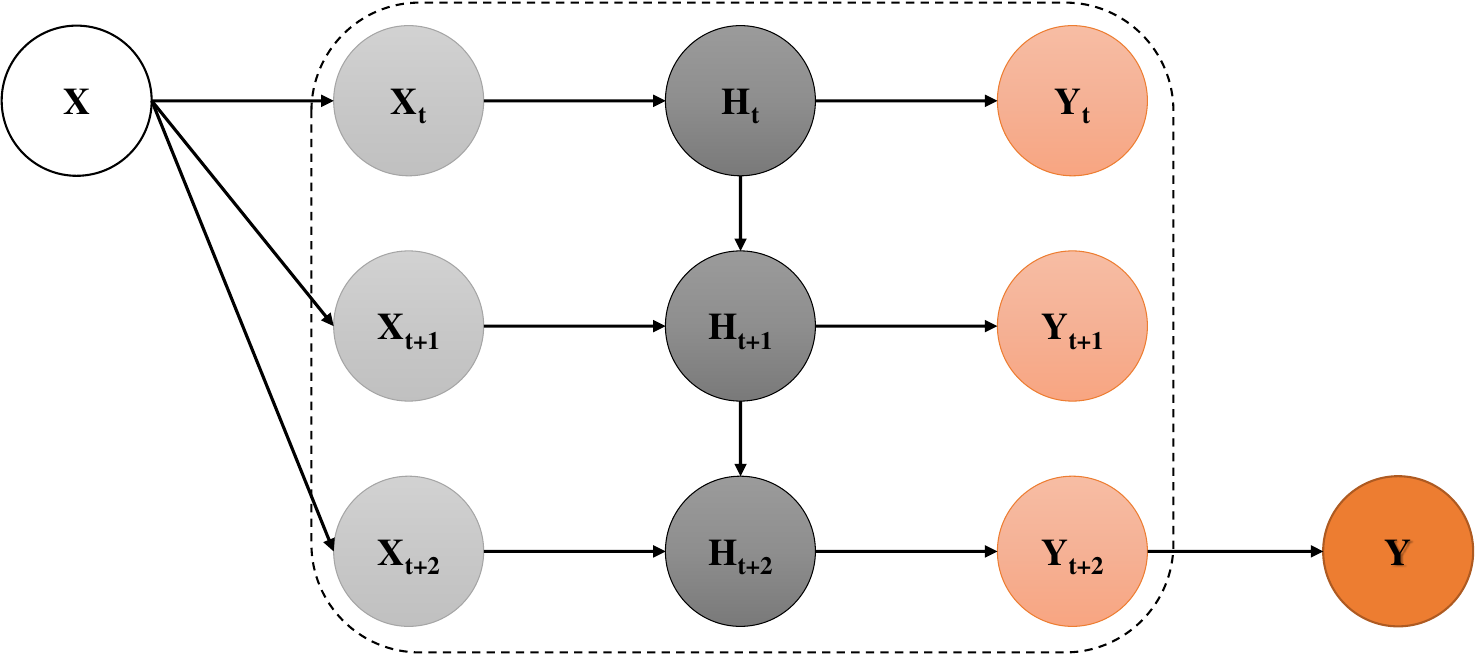}
\caption{Extended representation of a Recurrent Neural Network (RNN) for an example with an input sequence of length three, three hidden units, and a single output~\cite{shickel2017deep}.}
\label{fig:RNN}
\end{figure}

On the other side, in recent years, DL algorithms have become popular in many domains, due to their superior performance~\cite{liu2016action}. 
Since DL is based on the idea of the data representation, such techniques can automatically generate optimal features, starting from the raw input data, without any human intervention, making it possible to  identify the unknown patterns that otherwise would remain hidden or unknown~\cite{shickel2017deep}.
However, as already mentioned, DL models also present some limitation~\cite{marcus2018deep}:
\begin{itemize}
    \item Black-box models, interpretation is not easy and inherent,
    \item Require large datasets for training,
    \item High computational cost.
\end{itemize}
Because of such limitations, in some areas still CML methods are preferred, especially when the training dataset is quite small, or when fast training is a requirement. 
Some of the most common DL algorithms are: Convolutional Neural Network (CNN), Recurrent Neural Networks (RNNs), Long Short-Term Memory Networks (LSTMs), Gated Recurrent Unit (GRU), Stacked Autoencoders, Temporal Convolutional Network (TCN) and VAriational Autoencoders (VAE)~\cite{brownlee2016master}.
\begin{figure}[!t]
\centering 
\includegraphics[width=0.5\textwidth,page={20}]{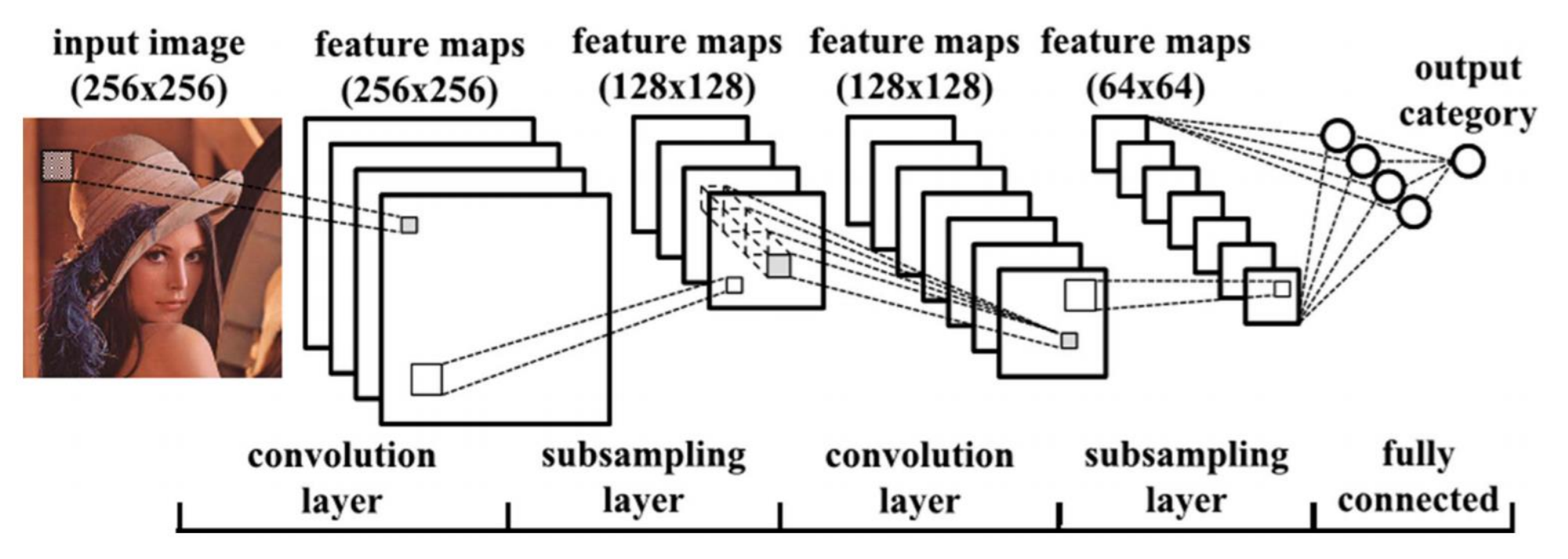}
\caption{Example of a Convolutional Neural Network (CNN) for image classification~\cite{shickel2017deep}.}
\label{fig:CNN}
\end{figure}
Nowadays, CNNs are a prevalent tool, especially in the image processing research community. CNN's impose local connectivity on the raw data extracting more important features by viewing the image as a collection of local pixel patches. Furthermore, a one-dimensional time series can also be viewed as a collection of local signal segments. Figure~\ref{fig:CNN} shows an example of CNN architecture with two convolutional layers, each followed by a pooling layer.
Instead, RNNs are a proper alternative when data is represented sequentially as time-series data and designed to deal with such long-range temporal dependencies. While one-dimensional sequences can be fed to a CNN, the resulting extracted features are shallow. Only closely localized relations between a few neighbors are factored into the feature representations. 
LSTM's are an RNN variant. Standard RNNs are comprised of interconnected hidden units, each unit in a Gated RNN is replaced by a special cell that contains an internal recurrence loop and a system of gates that controls the flow of information. Figure~\ref{fig:RNN} shows an RNNs that operates by sequentially updating a hidden state $H_t$ based not only on the activation of the current input $X_t$ at time $t$, but also on the previous hidden state $H_{t-1}$, updated by  $X_{t-1}$, $H_{t-2}$. The final hidden state after processing an entire sequence contains information from all its previous elements.
LSTM and GRU models are successful RNN variants, also known as Gated RNNs. Basic RNNs are comprised of interconnected hidden units. Each unit in a Gated RNN is substituted by a cell that includes an internal recurrence loop and a system of gates that manages the information flow. Gated RNNs have shown advantages in modeling sequential dependencies in long-term time-series~\cite{shickel2017deep}.

\subsection{Sensors}
Sensors and wearable devices surround us in our daily life. 
The most common types of sensors used in activity recognition are accelerometers, mainly due to their small size and low cost.  Figure~\ref{fig:five} illustrates the prevalence of accelerometer sensors used in HAR. In many cases, accelerometers are used in conjunction with others sensors including gyroscopes, magnetometers, compasses, pressure sensors, body temperature sensors, electromyography, oximetry sensors, and electrocardiographs. 
Many other kinds of sensors have been used in different applications. For example, the Global Positioning System (GPS) sensors or WiFi are used to determine the user’s location~\cite{kim2010sensloc}, microphones and Bluetooth are used to analyze human interactions~\cite{lu2009soundsense}, and $CO_2$ sensors are employed to estimate the air quality~\cite{lee2012comon}. 
The size of these sensors are constantly decreasing, such that they are being integrated into clothes~\cite{zhan2015multi}, smart glasses~\cite{gummeson2014energy} and other wearable objects~\cite{leonov2013thermoelectric}. 
In more advanced applications, objects in the daily environment are enriched with Radio Frequency Identification (RFID) tags. The tags make it possible to infer the user’s in-house activities (e.g., preparing coffee, doing laundry, washing dishes)~\cite{wang2012hierarchical}.\\

These sensors are becoming more and more prevalent in our daily life~\cite{rault2017survey}.
Shipments of wearable devices, including smartwatches, basic watches, and wrist bands, reached 34.2 million units during the second part of 2019, up 28.8\% year over year~\cite{ships_w}. 
Companies as Xiaomi, Apple, Huawei, Fitbit, or Samsung are pushing forward with new products capturing 65.7\% of the market, an almost 12\% more than 2018.~\cite{ships_w}.
Smart devices lend themselves to increasingly complex innovations in sensing and actuation. For example, when acceleration and inertial sensors are available, HAR algorithms can be implemented. Furthermore, by including additional electronic modules, such as Bluetooth Low Energy (BLE) and Wireless Local Area Network (WLAN) antennas or GPS, wearable devices can be used for real-time alerting and determining location to report risky situations and identify activity~\cite{demrozi2019indoor}. In addition to smartphones and smartwatches, other types of data collection and sensing systems with communication capabilities are adding to the Internet of Things (IoT).

\subsection{Sensor Fusion Techniques}
Each type of sensor provides benefits and disadvantages. 
For example, an accelerometer can measure acceleration, but cannot accurately evaluate velocity or positional changes. 
Similarly, the gyroscope can detect angular velocities, and the magnetometer can measure the magnetic field value. 
However, most sensors can easily be deceived by environmental noise, hardware  noise, or external inputs, resulting in imprecision and uncertainty. 
Sensor fusion techniques address these limitations by combining input from various sensors. 
The use of multiple sources (heterogeneous or homogeneous) combined with data fusion techniques provides several advantages, including 1) noise reduction, 2) reduced uncertainty, 3) increased robustness of the fusion phase, 4) robustness to interference, and 5) integration of prior knowledge of the perceived signals~\cite{uddin2020body,nweke2018analysis}. 
Generally, as the number of sensors increases, the fusion step becomes more challenging.
The most common sensor fusion methods are typically based on Bayesian estimation, Kalman Filters, and Particle Filtering techniques~\cite{zhu2019human}.
Nowadays, it is possible to implement these techniques directly at the hardware level inside the sensing modules, standardizing the application input and simplifying application development, maintenance, and extensibility. 
In the future, the use of sensor fusion techniques will span a wide range of applications~\cite{chen2017survey}.
Sensor fusion techniques address these limitations by combining the input from various sensors. 
The use of multiple sources (heterogeneous or homogeneous) combined with data fusion techniques provides several advantages, including 1) noise reduction, 2) lower uncertainty, 3) higher robustness, 4) robustness to interference, 5) integration of prior knowledge of the perceived signals~\cite{uddin2020body,nweke2018analysis}.
Generally, the more the number of sensors, the more challenging is the fusion step. 
The most common sensor fusion methods are typically based on Bayesian, Kalman Filter and Particle Filtering techniques~\cite{zhu2019human}.
Furthermore, nowadays, these techniques are directly imprinted at the hardware level inside the sensing modules, standardizing the application input and simplifying application development, maintenance, and extensibility.
In the future, the use of sensor fusion techniques will span a wide range of applications, given the specific functionality of each sensor and the need to obtain accurate and robust estimations~\cite{chen2017survey}.
\section{Human Activity}\label{sec:humact}
The definition of Activities of Daily Life (ADL's) is broad. ADL’s are the activities that we perform daily, such as eating, bathing, dressing, working, homemaking, enjoying leisure and all of these activities involving physical movement. Our review of HAR scientific literature presents an overview of the most studied ADL’s.

Among all ADL’s, the most popular activities in HAR research are walking, running, standing, sitting, walking upstairs and walking downstairs. However, other type of activities have been explored in the past few years, including complex activities, such as the different phases of cooking~\cite{lv2018bi}, house cleaning~\cite{liu2016action,liu2015action2activity,lv2018bi,arif2015physical}, driving~\cite{bhattacharya2016smart,bhat2018online,yao2017deepsense,siirtola2016user}, smoking~\cite{anazco2018smoking}, swimming~\cite{brunner2019swimming}, or biking~\cite{stisen2015smart,yao2017deepsense,san2018robust,abdallah2015adaptive}. 
Several studies focus on activities performed on specific locations, such as sitting on the ground, lying on bed~\cite{attal2015physical,wu2016mixed,wang2016recognition}, walking/standing in the elevator~\cite{wang2016recognition,badawi2018multimodal,zhou2019smartphone,civitarese2019context}, walking/running on a treadmill, walking in a parking lot, exercising on a stepper~\cite{wang2016recognition}, or exercising on a cross trainer~\cite{wang2016recognition,margarito2015user}. 
Other detailed movement recognition involves specific movements of the arms, such as carrying/reaching an object, releasing it, frontal elevation, and other activities that people can perform in relation to other objects~\cite{subasi2018sensor,ha2016convolutional,subasi2018iot}. 
A major area of HAR research involves the aging of population and the increasing of the number of people with physical and cognitive function impairments. Many HAR models are being used to help users recognize and avoid risky situations, such as falls in elderly people~\cite{masum2018human,ding2018energy,chen2015deep,micucci2017unimib,nait2018deep,sukor2018activity,tian2018adaptive}  or Freezing of Gait (FoG) in Parkinson’s disease~\cite{demrozi2019towards}. 
Furthermore, activity tracking devices are becoming very popular for monitoring ADLs. Those devices are able to approximate physiological and physical parameters such as heart rate, blood pressure, steps, level changes, and calories consumed. 
Advanced devices can recognize sleeping and the neurological stages of sleep (i.e., cycling through nREM (stages 1-4) and REM)~\cite{chetty2015smart}; furthermore, all the stored information can be used as input to HAR algorithms.
\section{Data Source Devices In HAR}\label{sec:device}
\begin{figure}[!th]
\centering
\minipage{0.45\textwidth}
\includegraphics[width=\linewidth,page={5}]{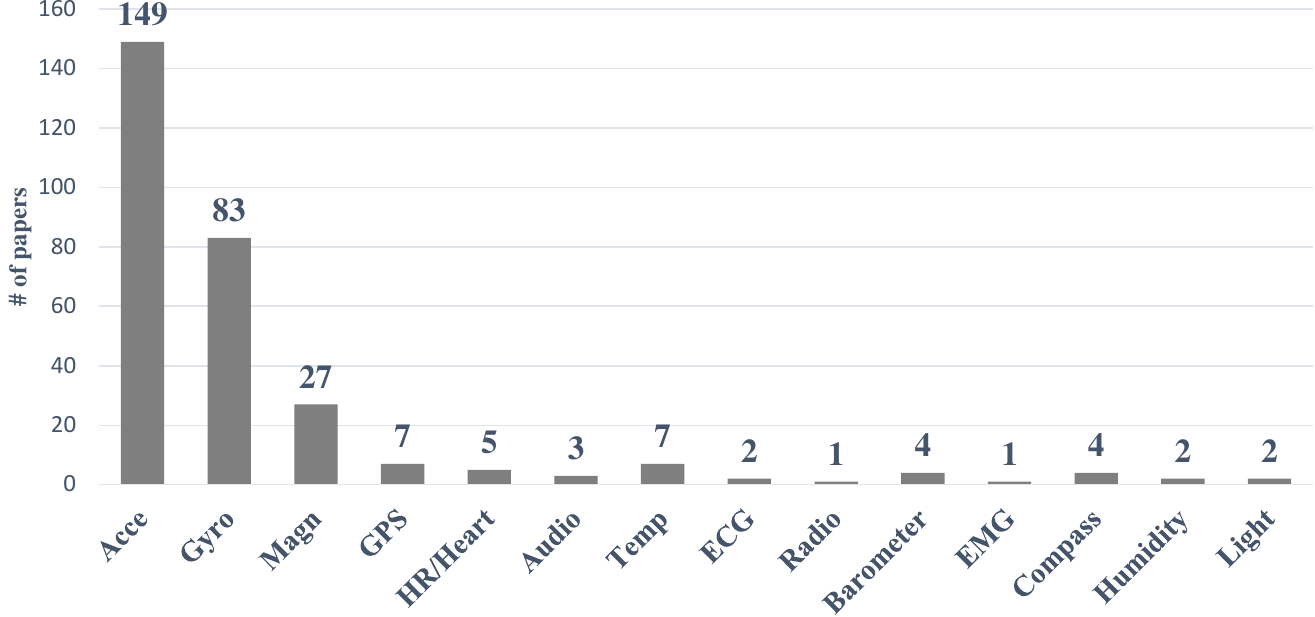} \\ \centering(a)
\endminipage\hfill
\minipage{0.45\textwidth}
\includegraphics[width=\linewidth,page={6}]{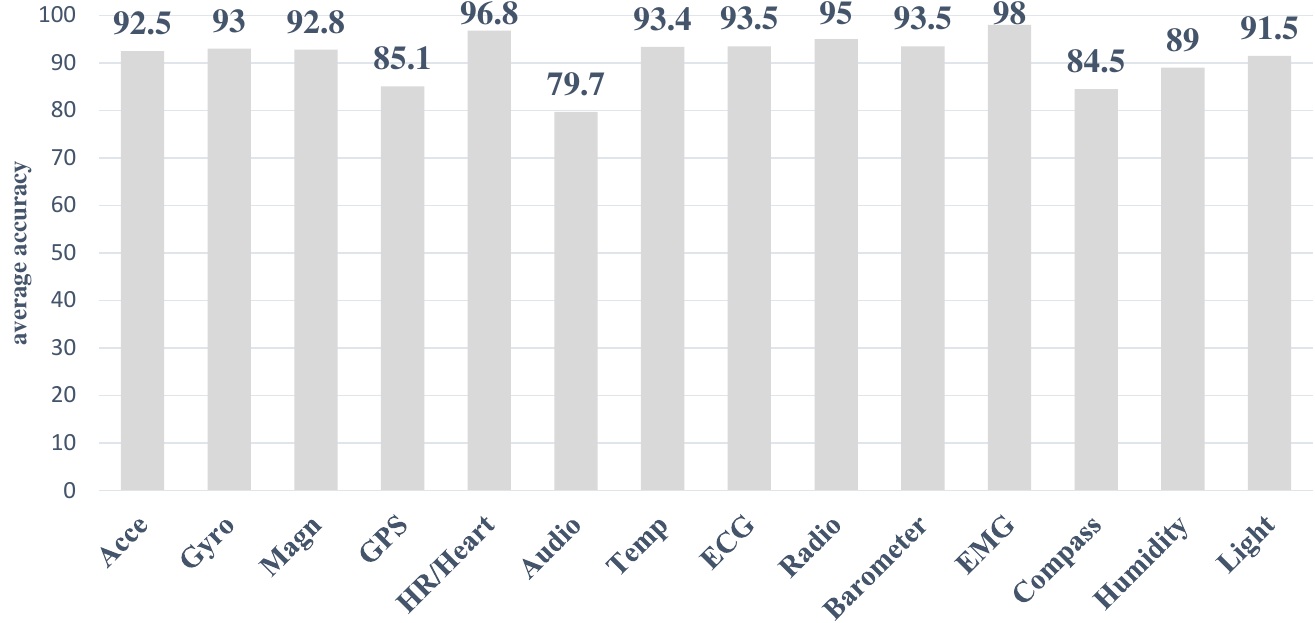} \\ \centering(b)
\endminipage\hfill 
\caption{(a) Distribution of published papers in HAR research area categorized by the sensor data source, and, (b) average activity recognition accuracy obtained from the papers using such sensors.}\label{fig:five}
\end{figure}
The first step of the HAR workflow includes identification of the data source sensor/device to be used, and, as shown in Figure~\ref{fig:five}.(a), small, low-cost and non-invasive sensors such as accelerometers, gyroscopes, and magnetometers are the most commonly used and appropriate sensors in HAR.
As depicted in Figure~\ref{fig:five}.(a), 149 papers used accelerometers, 83 used gyroscopes in addition to accelerometers, and 27 used a magnetometer in addition to the accelerometer. 
Therefore, all the selected papers use at least one accelerometer or at least one accelerometer in combination with other sensors. 
Furthermore, Figure~\ref{fig:five}.(b) shows the average activity recognition accuracy obtained form combination of such device.

Table~\ref{tab:two} and Table~\ref{tab:three} respectively show the sensor/device type and provide references to the papers using such sensors/device. Besides, Table~\ref{tab:two} and Table~\ref{tab:three} show in Columns Three to Five, the average number of recognized activities, average number of tested datasets and the average number of testing subject. 
These tables illustrate the importance of sensors like accelerometer, gyroscope, and magnetometer. However, other type of sensors as environmental sensors (temperature~\cite{kanjo2019deep,liu2015action2activity,subasi2018sensor,masum2018human,de2015multimodal,yang2015deep,hammerla2016deep}, humidity~\cite{masum2018human,de2015multimodal}, light~\cite{liu2015action2activity,lawal2019deep}, Passive Infrared Sensor (PIR)~\cite{yang2015deep}), radio signals (WiFi and Bluetooth~\cite{bhattacharya2016smart,de2015multimodal,uddin2020body}), medical equipment ( \ac{ECG}~\cite{ha2016convolutional,uddin2020body},  \ac{EMG}~\cite{badawi2018multimodal}) or other type of build in sensors (GPS~\cite{kanjo2019deep,de2015multimodal,lawal2019deep,abdallah2015adaptive,vaizman2017recognizing,cruciani2018automatic,polu2018human}, compass~\cite{vaizman2017recognizing,polu2018human}, heart rate~\cite{hammerla2016deep,rodriguez2017iot,balli2019human,manjarres2018human}, barometer~\cite{brunner2019swimming,zhou2019smartphone,ding2018energy}, stretch~\cite{bhat2018online,nk2019sensor}, audio~\cite{bhattacharya2016smart,lawal2019deep,vaizman2017recognizing,nguyen2018dealing}) are common in \ac{HAR}.
\begin{figure}[!th]
\centering
\minipage{0.35\textwidth}
\includegraphics[width=\linewidth,page={9}]{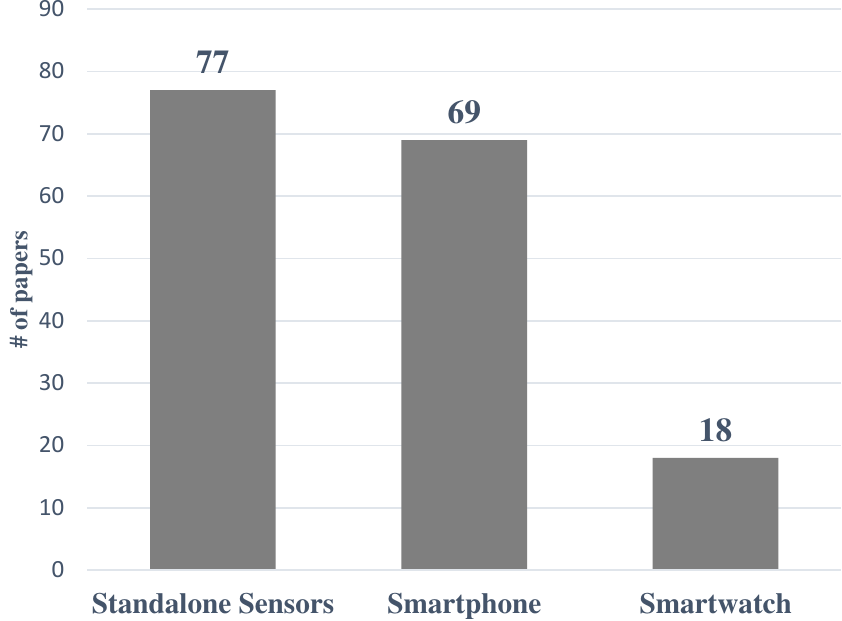} \\ \centering(a)
\endminipage\hfill
\minipage{0.35\textwidth}
\includegraphics[width=\linewidth,page={10}]{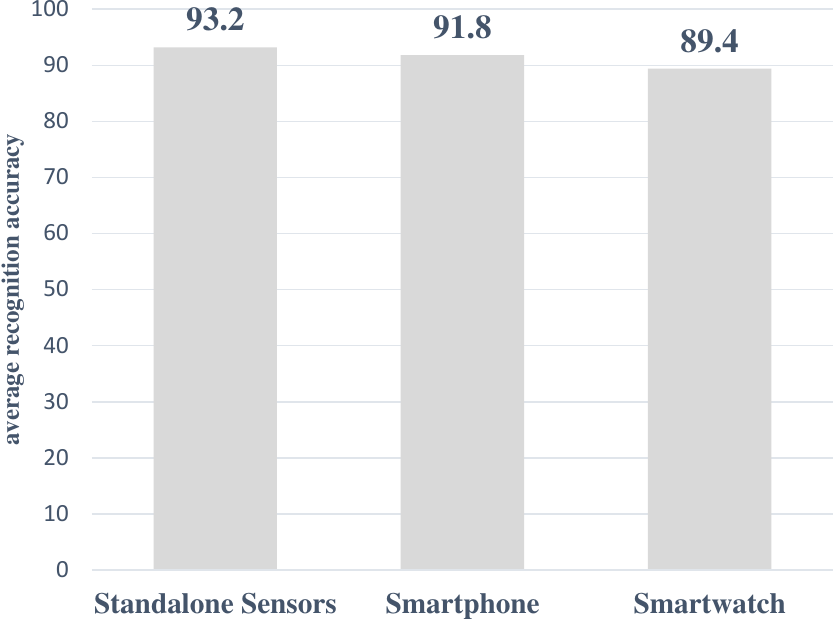} \\ \centering(b)
\endminipage\hfill 
\caption{(a) Distribution of published papers in HAR research area categorized by the device data source, and (b) Average activity recognition accuracy obtained by the identified devices.}\label{fig:six}
\end{figure}

\begin{table*}[!ht]
\caption{Sensor based paper categorization}
\label{tab:two}
\centering
\resizebox{0.95\textwidth}{!}{\begin{tabular}{c|p{9cm}|c|c|c}
\hline
Source Sensor           &
Article Reference       & 
\makecell{Average\\ \# Activities} & 
\makecell{Average\\ \# Datasets}   &
\makecell{Average\\ \# Subjects}   \\
\hline
Accelerometer&
\cite{ronao2016human,hammerla2016deep,yang2015deep,jiang2015human,hassan2018robust,chen2015deep,ordonez2016deep,alsheikh2016deep,ignatov2018real,alsheikh2016mobile,ravi2016deep,zheng2018comparison,guan2017ensembles,lawal2019deep,grzeszick2017deep,chen2017robust,chen2018novel,neverova2016learning,xu2019innohar,sathyanarayana2016sleep,ha2016convolutional,nait2018deep,shoaib2016complex,liu2016action,micucci2017unimib,civitarese2019context,rodriguez2017iot,suto2018efficiency,wang2018deep,nguyen2019wearable,li2019applying,subasi2018iot,liu2017wearable,ding2019empirical,nardi2019human,machado2015human,wang2016comparative,almaslukh2017effective,ponce2016novel,li2018comparison,jordao2018novel,badawi2018multimodal,nk2019sensor,masum2018human,nweke2018analysis,bulbul2018human,cao2018gchar,inoue2018deep,manjarres2018human,milenkoski2018real,cruciani2018automatic,lago2018improving,san2018robust,zhu2019human,mimouna2018human,kwon2018adding,bhat2018online,siirtola2019user,willetts2018semi,de2018human,nguyen2018dealing,balli2019human,sukor2018activity,chung2019sensor,choi2018temporal,zebin2018human,hossain2018improving,xu2018human,polu2018human,espinilla2018human,zhu2019efficient,su2018activity,rosati2018comparison,lv2018bi,subasi2018sensor,8727452,zhou2019smartphone,uddin2020body,nutter2018design,tian2018adaptive,voicu2019human,kanjo2019deep,almaslukh2018robust,wang2019attention,vaughn2018activity,malaise2018activity,zhao2018deep,ding2018energy,sun2018sequential,niu2018extreme,attal2015physical,reyes2016transition,lu2017towards,catal2015use,wannenburg2016physical,capela2015feature,zebin2016human,khalifa2017harke,ignatov2016human,wang2016triaxial,stisen2015smart,paul2015effective,capela2016evaluation,zhang2015recognizing,zubair2016human,heng2016human,torres2015robust,khan2015beyond,de2015multimodal,liu2015sensor,zheng2015human,tao2016multicolumn,akhavian2016smartphone,suarez2015improved,wu2016mixed,shen2016motion,chen2016lstm,chen2016lstm,zainudin2015activity,vavoulas2016mobiact,san2016feature,panwar2017cnn,liu2015action2activity,margarito2015user,zhang2015human,davis2016activity,lubina2015artificial,yin2015human,weiss2016smartwatch,zhu2015smartphone,abdallah2015adaptive,lee2016energy,arif2015physical,capela2015improving,mannini2017activity,wang2016recognition,damavsevivcius2016human,vaizman2017recognizing} &12.84&1.32&45.38\\ \hline
Gyroscope    &
\cite{ronao2016human,hammerla2016deep,yang2015deep,jiang2015human,hassan2018robust,ravi2016deep,guan2017ensembles,lawal2019deep,chen2018novel,neverova2016learning,xu2019innohar,ha2016convolutional,shoaib2016complex,liu2016action,micucci2017unimib,civitarese2019context,suto2018efficiency,wang2018deep,nguyen2019wearable,li2019applying,subasi2018iot,liu2017wearable,ding2019empirical,nardi2019human,wang2016comparative,almaslukh2017effective,ponce2016novel,li2018comparison,badawi2018multimodal,masum2018human,nweke2018analysis,bulbul2018human,cao2018gchar,inoue2018deep,zhu2019human,balli2019human,chung2019sensor,choi2018temporal,zebin2018human,polu2018human,espinilla2018human,huang2019tse,zhu2019efficient,su2018activity,rosati2018comparison,subasi2018sensor,zhou2019smartphone,uddin2020body,nutter2018design,voicu2019human,kanjo2019deep,wang2019attention,vaughn2018activity,malaise2018activity,zhao2018deep,ding2018energy,niu2018extreme,attal2015physical,reyes2016transition,catal2015use,capela2015feature,zebin2016human,ignatov2016human,paul2015effective,capela2016evaluation,zhang2015recognizing,de2015multimodal,liu2015sensor,akhavian2016smartphone,suarez2015improved,shen2016motion,san2016feature,liu2015action2activity,davis2016activity,yin2015human,weiss2016smartwatch,zhu2015smartphone,abdallah2015adaptive,wang2016recognition,damavsevivcius2016human,vaizman2017recognizing}&14.22&1.33&43.74\\ \hline 
Magnetometer & 
\cite{hammerla2016deep,lawal2019deep,neverova2016learning,xu2019innohar,ha2016convolutional,micucci2017unimib,civitarese2019context,nguyen2019wearable,ding2019empirical,ponce2016novel,li2018comparison,siirtola2019user,de2018human,chung2019sensor,zhu2019efficient,subasi2018sensor,zhou2019smartphone,uddin2020body,voicu2019human,kanjo2019deep,wang2019attention,zhao2018deep,attal2015physical,zhang2015recognizing,liu2015action2activity,wang2016recognition}&17.45&1.44&84.25 \\\hline
Other        &
\cite{hammerla2016deep,yang2015deep,guan2017ensembles,lawal2019deep,ha2016convolutional,civitarese2019context,rodriguez2017iot,li2018comparison,badawi2018multimodal,nk2019sensor,masum2018human,manjarres2018human,cruciani2018automatic,bhat2018online,nguyen2018dealing,balli2019human,polu2018human,subasi2018sensor,zhou2019smartphone,uddin2020body,kanjo2019deep,zhao2018deep,ding2018energy,niu2018extreme,paul2015effective,de2015multimodal,liu2015action2activity,yin2015human,abdallah2015adaptive,vaizman2017recognizing}&21.09&2.09&29.45\\
\hline
\hline
\multicolumn{5}{c}{Other=\{temperature, humidity, light, presence, WiFi, Bluetooth, ECG,  EMG, GPS, compass, heart rate, barometer, strech, audio\} }
\end{tabular}}
\end{table*}
In addition to the direct measurements that such sensors provide, the indirect usage of the measurements in form of smart metrics is promising (e.g., energy harvesting of the system~\cite{khalifa2015pervasive} or the \ac{RSSI}~\cite{uddin2020body}) in order to recognize human activity related to direct measurements from the body or environmental variations. 
Furthermore, the importance of smartphones and smartwatches in HAR is increasingly clear, mainly due to their explosion among consumers and given that these devices currently contain many of the aforementioned sensors.
\begin{table*}[!ht]
\caption{Device based paper categorization}
\label{tab:three}
\centering
\resizebox{0.95\textwidth}{!}{\begin{tabular}{c|p{9cm}|c|c|c}
\hline
Source Device           &
Article Reference       & 
\makecell{Average\\ \# Activities} & 
\makecell{Average\\ \# Datasets}   &
\makecell{Average\\ \# Subjects}   \\
\hline
Standalone& 
\cite{hammerla2016deep,yang2015deep,ravi2016deep,jiang2015human,ordonez2016deep,alsheikh2016deep,ignatov2018real,zheng2018comparison,guan2017ensembles,grzeszick2017deep,xu2019innohar,sathyanarayana2016sleep,ha2016convolutional,nait2018deep,liu2016action,rodriguez2017iot,suto2018efficiency,wang2018deep,nguyen2019wearable,subasi2018iot,liu2017wearable,ding2019empirical,nardi2019human,machado2015human,ponce2016novel,li2018comparison,jordao2018novel,badawi2018multimodal,nk2019sensor,nweke2018analysis,manjarres2018human,lago2018improving,san2018robust,zhu2019human,mimouna2018human,kwon2018adding,bhat2018online,de2018human,nguyen2018dealing,sukor2018activity,chung2019sensor,choi2018temporal,zebin2018human,hossain2018improving,su2018activity,rosati2018comparison,subasi2018sensor,8727452,uddin2020body,tian2018adaptive,voicu2019human,malaise2018activity,sun2018sequential,attal2015physical,catal2015use,wannenburg2016physical,zebin2016human,ignatov2016human,wang2016triaxial,stisen2015smart,paul2015effective,zhang2015recognizing,zubair2016human,khan2015beyond,zheng2015human,wu2016mixed,liu2015action2activity,lubina2015artificial,mannini2017activity,wang2016recognition}&15.63&1.48&26.9\\ \hline
Smartphone& 
\cite{ronao2016human,hassan2018robust,chen2015deep,alsheikh2016mobile,ravi2016deep,lawal2019deep,chen2017robust,chen2018novel,neverova2016learning,shoaib2016complex,micucci2017unimib,civitarese2019context,suto2018efficiency,wang2016comparative,almaslukh2017effective,masum2018human,bulbul2018human,cao2018gchar,inoue2018deep,milenkoski2018real,cruciani2018automatic,siirtola2019user,xu2018human,polu2018human,espinilla2018human,huang2019tse,zhu2019efficient,lv2018bi,zhou2019smartphone,kanjo2019deep,almaslukh2018robust,wang2019attention,vaughn2018activity,zhao2018deep,ding2018energy,niu2018extreme,reyes2016transition,lu2017towards,capela2015feature,khalifa2017harke,capela2016evaluation,heng2016human,torres2015robust,de2015multimodal,liu2015sensor,tao2016multicolumn,akhavian2016smartphone,suarez2015improved,shen2016motion,chen2016lstm,chen2016lstm,zainudin2015activity,vavoulas2016mobiact,san2016feature,panwar2017cnn,zhang2015human,davis2016activity,yin2015human,weiss2016smartwatch,zhu2015smartphone,abdallah2015adaptive,lee2016energy,arif2015physical,capela2015improving,damavsevivcius2016human,vaizman2017recognizing}&10.55&1.18&65.59\\ \hline
Smartwatch& 
\cite{bhattacharya2016smart,lawal2019deep,civitarese2019context,li2019applying,willetts2018semi,balli2019human,lv2018bi,nutter2018design,kanjo2019deep,almaslukh2018robust,zhao2018deep,niu2018extreme,de2015multimodal,margarito2015user,weiss2016smartwatch,abdallah2015adaptive}&17.4&1.28&32\\ \hline
Other&
\cite{hammerla2016deep,yang2015deep,guan2017ensembles,nk2019sensor,masum2018human,nguyen2018dealing,subasi2018sensor,zhao2018deep,niu2018extreme,de2015multimodal,vaizman2017recognizing}&7&1&22\\
\hline
\hline
\multicolumn{5}{c}{Other =\{ temperature, humidity, light, PIR, WiFi, Bluetooth, heart rate, barometer, strech, audio, medical devices \} }
\end{tabular}}
\end{table*}
Finally, as shown in Figure~\ref{fig:six}.(a), among all the reviewed published papers, the proposed HAR methods are based mostly on standalone devices. However, the total number of smartphone- and smartwatch-based methods are higher than those based on standalone devices. 
Figure~\ref{fig:six}.(b) shows that in terms of recognition accuracy methodologies based on smartphone and smartwatch devices are in line with those obtained from standalone devices. Moreover, smartphones and smartwatches~\cite{kheirkhahan2019smartwatch}, unlike standalone devices, provide computational capabilities that make it possible to directly execute HAR models on the wearable device, and in many cases, they have a very high cost (e.g., devices used in the medical field).
\section{Data}\label{sec:data}
The second step of the HAR workflow regards the collected data type. Such data can mainly be categorized as follows.
\begin{itemize}
    \item Inertial sensors data such as accelerometers, gyroscopes, magnetometer, or compass,
    \item Physiological sensors data such as \ac{ECG}, \ac{EMG}, Heart Rate, or blood pressure,
    \item Environmental sensors data such as temperature, pressure, $CO_2$, humidity, or PIR.
\end{itemize}

\subsection{Inertial sensors data}
Accelerometer, gyroscope, and magnetometer sensors with a maximum of nine degrees of freedom are commercially available at a very low cost. Besides, acceleration and angular velocity are the most common data used to characterize human activity. This is reinforced by what we described in the previous section, given that accelerometers and the gyroscopes are the most widely used devices in HAR. Such inertial sensors are widely used in clinical and healthcare applications.~\cite{tamura2014wearable} 
\begin{table*}[!ht]
\caption{Data source used in HAR paper.}
\label{tab:four}
\centering
\resizebox{0.9\textwidth}{!}{\begin{tabular}{c|p{8cm}|c|c|c}
\hline
Data Type&
Article Reference       & 
\makecell{Average\\ \# Activities} & 
\makecell{Average\\ \# Datasets}   &
\makecell{Average\\ \# Subjects}   \\
\hline
Inertial     &
\cite{ronao2016human,hammerla2016deep,yang2015deep,jiang2015human,hassan2018robust,chen2015deep,ordonez2016deep,alsheikh2016deep,ignatov2018real,alsheikh2016mobile,ravi2016deep,zheng2018comparison,guan2017ensembles,lawal2019deep,grzeszick2017deep,chen2017robust,chen2018novel,neverova2016learning,xu2019innohar,sathyanarayana2016sleep,ha2016convolutional,nait2018deep,shoaib2016complex,liu2016action,micucci2017unimib,civitarese2019context,rodriguez2017iot,suto2018efficiency,wang2018deep,nguyen2019wearable,li2019applying,subasi2018iot,liu2017wearable,ding2019empirical,nardi2019human,machado2015human,wang2016comparative,almaslukh2017effective,ponce2016novel,li2018comparison,jordao2018novel,badawi2018multimodal,nk2019sensor,masum2018human,nweke2018analysis,bulbul2018human,cao2018gchar,inoue2018deep,manjarres2018human,milenkoski2018real,cruciani2018automatic,lago2018improving,san2018robust,zhu2019human,mimouna2018human,kwon2018adding,bhat2018online,siirtola2019user,willetts2018semi,de2018human,nguyen2018dealing,balli2019human,sukor2018activity,chung2019sensor,choi2018temporal,zebin2018human,hossain2018improving,xu2018human,polu2018human,espinilla2018human,zhu2019efficient,su2018activity,rosati2018comparison,lv2018bi,subasi2018sensor,8727452,zhou2019smartphone,uddin2020body,nutter2018design,tian2018adaptive,voicu2019human,kanjo2019deep,almaslukh2018robust,wang2019attention,vaughn2018activity,malaise2018activity,zhao2018deep,ding2018energy,sun2018sequential,niu2018extreme,attal2015physical,reyes2016transition,lu2017towards,catal2015use,wannenburg2016physical,capela2015feature,zebin2016human,khalifa2017harke,ignatov2016human,wang2016triaxial,stisen2015smart,paul2015effective,capela2016evaluation,zhang2015recognizing,zubair2016human,heng2016human,torres2015robust,khan2015beyond,de2015multimodal,liu2015sensor,zheng2015human,tao2016multicolumn,akhavian2016smartphone,suarez2015improved,wu2016mixed,shen2016motion,chen2016lstm,zainudin2015activity,vavoulas2016mobiact,san2016feature,panwar2017cnn,liu2015action2activity,margarito2015user,zhang2015human,davis2016activity,lubina2015artificial,yin2015human,weiss2016smartwatch,zhu2015smartphone,abdallah2015adaptive,lee2016energy,arif2015physical,capela2015improving,mannini2017activity,wang2016recognition,damavsevivcius2016human,vaizman2017recognizing,huang2019tse} &12.88&1.32&45.54\\ \hline 
Physiological&
\cite{ha2016convolutional,uddin2020body,badawi2018multimodal,manjarres2018human,balli2019human,rodriguez2017iot,hammerla2016deep,kanjo2019deep} &12.71&1.57&11\\ 
\hline
Environmental&
\cite{kanjo2019deep,subasi2018sensor,hammerla2016deep,masum2018human,de2015multimodal,liu2015action2activity,yang2015deep,zhou2019smartphone,ding2018energy,lawal2019deep,vaizman2017recognizing,nguyen2018dealing,khalifa2015pervasive} &20.76&1.47&19.59\\ \hline
\hline
\end{tabular}}
\end{table*}

\subsection{Physiological sensors Data}
Physiological sensors perceive physiological signals, which in contradiction with other sources of emotional knowledge (facial, gestures, and speech), providing essential advantages. Those signals are mostly involuntary and, as such, are quite insensitive to deception. 
They can be used to continuously measure the affective events.~\cite{feidakis2016review}
The most used physiological signals are brain electrical activity, heartbeat, muscle electrical activity, blood pressure, and skin conductance acquired by the following external data acquisition system: Electroencephalogram (EEG), Electrocardiogram (ECG), and Electromyography (EMG).

\subsection{Environmental sensors data}
The environmental data covers all the collection of data representing the state of the environment, including temperature, humidity, pressure, or brightness. However, measuring the status of the environment goes beyond environmental measures. It can also include more complex measures related to people and objects inside the environment. 
For example, recognizing the number of people inside the environment and their position or the actions performed on a certain object inside the environment could be useful in many application scenarios related to human assistance, healthcare, and service delivery.

\noindent Table~\ref{tab:four} shows the categorization of the revised articles based on the type of data, where Column One and Two show the data type and the reference to the articles using such data types. Columns Three to Five respectively show the average number of recognized activities, average number of tested datasets, and average number of testing subjects.
However, as we discussed earlier, the largest amount of data on daily life is collected via electronic devices, such as smartphones, smartwatches, activity trackers, smart thermostats and video cameras. 
As shown in Figure~\ref{fig:six}, the use of smart devices like smartphone and smartwatch is outnumbering the use of standalone devices. It should be noted that the standalone column identifies all those devices other than smartphones and smartwatches as for example, clinical and  dedicated instruments, such as Actigraph (Actigraph, Florida/USA), or Bioharness3 (RAE Systems by Honeywell, California/USA).
Furthermore, during the data collection step, sometimes activities are performed in a controlled manner (aka scrippted). That is because human movement patterns are very hard to recognize due to the large inter-subject and intra-subject variability. Such variability entails a considerable difficulty in developing a methodology that manages to generalize among all subjects. Also, the lack of data collected from a very large number of subjects does not help researchers find a solution to this problem.

\noindent With regard to such issue, Table~\ref{tab:five} shows some of the best known and open source datasets for HAR studies.
\begin{table*}[t!]
\centering
\caption{Publicly Available Datasets for HAR research}\label{tab:five}
\resizebox{0.95\textwidth}{!}{
\begin{tabular}{p{2.25cm} p{7cm} l p{4cm} l l}
\hline
\textbf{Dataset} & \textbf{Activities} & \textbf{\# Activities} & \textbf{Data Sources} & \textbf{\# Subjects} & \textbf{Citations} \\
\hline
\hline
\href{http://www.cis.fordham.edu/wisdm/dataset.php}{WISDM v1}\cite{kwapisz2011activity}&
walking, jogging, upstairs, downstairs, sitting, standing&
6&
Smartphone accelerometer(controlled environments)&
26&1939 \\
\hline
\href{https://archive.ics.uci.edu/ml/datasets/opportunity+activity+recognition}{Opportunity}]~\cite{roggen2010collecting}&
Start, groom, relax, prepare coffee, drink coffee, prepare sandwich, eat sandwich, cleanup, break, open and close: fridge, dishwasher, drawers, door 1, door 2, on/off lights, drink standing, drink sitting &
18&	23 body sensors
12 object sensors
21 ambient sensors&
4&367 \\
\hline
\href{https://archive.ics.uci.edu/ml/datasets/human+activity+recognition+using+smartphones}{UCI-HAR}~\cite{anguita2013public}&
walking, upstairs, downstairs, sitting, standing, laying&
6&	
Samsung Galaxy S II accelerometer, gyroscope&
30&635 \\
\hline
\href{http://sipi.usc.edu/had/}{USC-HAD}~\cite{zhang2012usc}&
walking: forward, left, right, upstairs, downstairs, running forward, jumping, sitting, standing, sleeping, elevator up, elevator down&
12&	
MotionNode accelerometer&
14&180 \\
\hline
\href{http://har-dataset.org/doku.php?id=wiki:dataset}{Skoda} ~\cite{zappi2007activity}&
write notes, open engine hood, close engine hood, check door gaps, open door, close door, open/close two doors, check trunk gap, open/close trunk, check steering wheel&
10&
20 accelerometers&
1&38 
\\
\hline
\href{https://archive.ics.uci.edu/ml/datasets/pamap2+physical+activity+monitoring}{PAPAM2}~\cite{reiss2012introducing}&
lying, sitting, standing, walking, running, cycling, nordic walking, watching TV, computer work, car driving, ascending stairs, descending stairs, vacuum cleaning, ironing, folding laundry, house cleaning, playing soccer, rope jumping&
18&
3 colibri wireless inertial measurement units (accelerometer, gyroscope, magnetometer)&
9&397 \\
\hline
\href{https://archive.ics.uci.edu/ml/datasets/Daphnet+Freezing+of+Gait}{Daphnet}~\cite{bachlin2009wearable}&
freeze (gait block), no freeze (any activity different from gait block)&
2&
3 accelerometers (ankle, upper leg, trunk)&
10&319 \\
\hline
\href{https://archive.ics.uci.edu/ml/datasets/MHEALTH+Dataset}{mHealth}~\cite{banos2014mhealthdroid}&
standing still, Sitting and relaxing, lying down, walking, climbing stairs, waist bends forward, frontal elevation of arms, knees bending (crouching), cycling, jogging, running, jump front and back&
12&	
chest (accelerometer, gyroscope, magnetometer, ECG) right wrist (accelerometer, gyroscope, magnetometer) and left ankle (accelerometer, gyroscope, magnetometer)&
10&120 \\
\hline
\href{https://archive.ics.uci.edu/ml/datasets/Heterogeneity+Activity+Recognition}{HHAR}~\cite{stisen2015smart}&
biking, sitting, standing, walking, stair up and stair down&
6&
accelerometer, gyroscope from 8 smartphone and 4 smartwatches &
9&204 \\
\hline
\href{http://www.cis.fordham.edu/wisdm/dataset.php}{WISDM v2}~\cite{kwapisz2011activity}&
walking, jogging, upstairs, downstairs, sitting, standing&
6&
Smartphone accelerometer (un-controlled environments)&
563&1939 \\
\hline
\href{https://archive.ics.uci.edu/ml/datasets/daily+and+sports+activities}{DSADS}~\cite{altun2010comparative}&
Sitting, standing, lying on back and on right side, ascending and descending stairs, standing in an elevator still and moving around in an elevator, walking in a parking lot, walking on a treadmill with a speed of 4 km/h (in flat and 15 deg inclined positions), running on a treadmill with a speed of 8 km/h, exercising on a stepper, exercising on a cross trainer, cycling on an exercise bike in horizontal and vertical positions, rowing, jumping, and playing basketball&
19&	
5 units on torso, right arm, left arm, right leg, left leg 9 sensors on each unit (x,y,z accelerometers, x,y,z gyroscopes, x,y,z magnetometers)&
8&394 \\
\hline
\href{https://archive.ics.uci.edu/ml/datasets/REALDISP+Activity+Recognition+Dataset}{REALDISP}~\cite{banos2012benchmark}&
\small{walking, jogging, running, jump up, jump front and back, jump sideways, jump leg/arms open/closed , jump rope, trunk twist (arms outstretched), trunk twist (elbows bent), waist bends forward, waist rotation, waist bends (reach foot with opposite hand), reach heels backwards, lateral bend (10 to the left + 10 to the right), lateral bend with arm up (10 to the left + 10 to the right), repetitive forward stretching, upper trunk and lower body opposite twist, lateral elevation of arms, frontal elevation of arms, frontal hand claps, frontal crossing of arms, shoulders high-amplitude rotation, shoulders low-amplitude rotation, arms inner rotation, knees (alternating) to the breast, heels (alternating) to the backside, knees bending (crouching), knees (alternating) bending forward, rotation on the knees, rowing, elliptical bike, cycling}	&
33&
accelerometer, gyroscope, magnetometer, 4D quaternions on 9 positions: left calf, left thigh, right calf, right thigh, back, left lower arm, left upper arm, right lower arm, right upper arm &
17&80 \\
\hline
\href{http://www.sal.disco.unimib.it/technologies/unimib-shar/}{UniMiB SHAR}~\cite{micucci2017unimib}&
standing up from laying, lying down from standing, standing up from sitting, running, sitting down, downstairs, upstairs, walking, jumping&
17&
Smartphone  accelerometer&
30&75 \\
\hline
\href{http://hamlyn.doc.ic.ac.uk/activemiles/}{ActiveMiles}~\cite{active_w}&
Activities of daily life&
7&	
Smartphone accelerometer and gyroscope in uncontrolled environments&
10&72 \\
\hline
\href{https://people.eecs.berkeley.edu/~yang/software/WAR/}{WARD}~\cite{yang2009distributed}&
stand, sit, lie down, walk forward, walk left-circle, walk right-circle, turn left, turn right, upstairs, downstairs, jog, jump, push wheelchair&
13&
5 motion sensors (accelerometer, gyroscope)
2 on the wrists, one on the waist, and 2 on the ankles &
20&194\\
\hline
\hline
\end{tabular}}
\end{table*}

Column One refers to the name and the article proposing the dataset. Column Two presents the activity labeled in the dataset, Column Three shows the number of activities. Column Four shows the number and type of the used sensing devices. Column Five and Column Six show the number of subjects from whom the data was collected and the number of citations that the dataset received by September 2019.
Such datasets are largely based on accelerometer, gyroscope, and magnetometer sensor data. Most of such sensors are embedded into smartphones and smartwatches, and the number of activities in these datasets ranges from two~\cite{bachlin2009wearable} to thirty-three~\cite{banos2012benchmark} (Table~\ref{tab:five}).
The most common studied activities are primary activities of daily life, such as walking, running, sitting, standing, walking upstairs, walking downstairs, and sleeping.

\subsection{Preprocessing and Feature Extraction}
The mentioned data sources generate time-series information identifying the status of the device. However, data is characterized by noise, which makes it difficult to be used in their raw state. The presence of noise is handled by preprocessing the raw data to eliminate this interference and prepare the data for being feed to the recognition models~\cite{shoaib2015survey}. 
The preprocessing is one of the most important phases in HAR and presents different noise management techniques, such as digital and statistical filters, data normalization, and feature extraction. 
The features extraction step explores basically tow domains: time, frequency and spectral domain. Time-domain features are the most used because they are cheaper than the frequency domain features because of the transform from time to frequency domain~\cite{shoaib2015survey}.
Since standard classification models are not suitable for raw data, this phase is anticipated by a segmentation step during which time-series sensor data is segmented before extracting features. 
Besides, many methodologies maintain an overlapping part between two consecutive segments. This part provides the model with knowledge of the previous context. Table~\ref{tab:tf_features} presents an overview on the most commonly used time and frequency domain features.

Table~\ref{tab:pre-process} presents a categorization of the reviewed papers based on the utilization of noise removal, time domain, and frequency domain features extraction techniques. Columns One to Four show: the machine learning category (CML or DL), if any noise removal technique is used, if time-domain or frequency-domain features were extracted. Column Five contains the references to the papers, and Column Six, the number of papers using such configuration. Finally, Columns Seven and Eight show the average number of used features and the average activity recognition accuracy. Concerning the CML-based models, as shown, most of the reviewed articles (Tab.~\ref{tab:pre-process}, row 7) make use of both time and frequency domain features, and the raw data was initially pre-processed with noise removal techniques. Instead, Tab.~\ref{tab:pre-process}, row 3 shows articles that use time and frequency domain features without applying any noise removal technique. However, other methodologies (Tab.~\ref{tab:pre-process}, rows 8 and 9) do not make use of any features extraction technique, and in some cases, the presence of noise is not considered, as in~\cite{rodriguez2017iot,khan2015beyond,liu2015action2activity,abdallah2015adaptive,liu2016action,nguyen2018dealing}.
In~\cite{liu2016action,liu2015action2activity} and \cite{khan2015beyond}, the methodologies are based on the mining of temporal patterns and their symbolic representation, or as in~\cite{abdallah2015adaptive} were, authors make use of clustering technique, discriminating between different human activities. About the results obtained in terms of accuracy, the methodologies that make use of noise removal methods and feature extraction in the time and frequency domain show promising results as also shown by the number of methodologies that make use of this configuration. 

Furthermore, concerning the DL-based methodologies, since DL networks perform automatic feature extraction without human intervention, unlike traditional machine-learning algorithms, the majority of them do not make use of any Noise Removal and Feature Extraction step as shown in Tab.~\ref{tab:pre-process}, rows 10 and 14. The achieved average accuracy, among all these 34 articles, was 93\%. Besides, other DL-based articles do make use of time-domain (Tab.~\ref{tab:pre-process}, row 12)  features, frequency domain (Tab.~\ref{tab:pre-process}, row 11) and both time and frequency domain (Tab.~\ref{tab:pre-process}, rows 13 and 14) features.
DL-based models eliminate the latency due to the need to process data with the above techniques. However, such models require a more considerable amount of data than ML models and longer training times. 

Concerning the Noise Removal step, 48 CML-based articles and 12 DL-based articles make use of different noise removal techniques. Among all such techniques the most used ones are: z-normalization~\cite{nardi2019human,margarito2015user}, min-max~\cite{bulbul2018human,wu2016mixed}, and linear interpolation~\cite{ordonez2016deep,xu2019innohar} are the most used normalization steps, preceded by a filtering step based on the application of outlier detection~\cite{li2019applying,wang2016triaxial,wu2016mixed}, Butterworth~\cite{hassan2018robust,almaslukh2017effective,zhao2018deep,micucci2017unimib,li2019applying,bulbul2018human,cao2018gchar,reyes2016transition,suarez2015improved,mannini2017activity}, median~\cite{hassan2018robust,nutter2018design,civitarese2019context,li2019applying,bulbul2018human,mimouna2018human,reyes2016transition,davis2016activity,lubina2015artificial}, high-pass~\cite{li2019applying,cao2018gchar,manjarres2018human,cruciani2018automatic,torres2015robust,akhavian2016smartphone,pham2015mobirar}, or statistical~\cite{zhu2019human} filters.

\begin{table*}[!ht]
\caption{Most used Time and Frequency domain features}
\label{tab:tf_features}
\centering
\resizebox{\textwidth}{!}{\begin{tabular}{ p{9cm} || p{9cm} }
\hline
\hline
Time Domain Features       & 
Frequency Domain Features  \\
\hline
\hline
1) maximum, 
2) minimum, 
3) mean, 
4) standard deviation,
5) root mean square, 
6) range, 
7) median, 
8) skewness, 
9) kurtosis, 
10) time-weighted variance, 
11) interquartile range, 
12) empirical cumulative density function, 
13) percentiles (10, 25, 75, and 90), 
14) sum of values above or below percentile (10, 25, 75, and 90), 
15) square sum of values above or below percentile (10, 25, 75, and 90), 
16) number of crossings above or below percentile (10, 25, 75, and 90), 
17) mean amplitude deviation,
18) mean power deviation, 
19) signal magnitude area, 
20) signal vector magnitude, 
21) covariance, 
22) simple moving average of sum of range of a signal, 
23) sum of range of a signal, 
24) sum of standard deviation of a signal, 
25) maximum slope of simple moving average of sum of variances of a signal, 
26) autoregression.
& 
1) fast fourier transform (FFT) coefficients, 
2) discrete fourier transform (DFT), 
3) discrete wavelet transform (DWT), 
4) first dominant frequency, 
5) ratio between the power at the dominant frequency and the total power, 
6) ratio between the power at frequencies higher than 3.5 Hz and the total power, 
7) two signal fragmentation features, 
8) DC component in FFT spectrum, 
10) energy spectrum, 
11) entropy spectrum, 
12) sum of the wavelet coefficients, 
13) squared sum of the wavelet coefficients and energy of the wavelet coefficients, 
14) auto-correlation, 
15) mean-crossing rate, 
16) spectral entropy, 
17) spectral energy, 
18) wavelet entropy values, 
19) mean frequency, 
20) energy band \\
\hline
\hline
\end{tabular}}
\end{table*}

\begin{table*}[!bht]
\caption{Preprocessing and Feature extraction on the reviewed papers.}
\label{tab:pre-process}
\centering
\resizebox{0.95\textwidth}{!}{\begin{tabular}{c || c c c || p{7cm} c c c}
\hline
\hline
\makecell{ML\\ Model}                       &
\makecell{Noise\\ Removal}                  &
\makecell{Time\\ Domain\\ Features}         &    
\makecell{Frequency\\ Domain\\ Features}    &
\makecell{Papers\\ Reference}               &
\makecell{\# of\\Papers}                    &
\makecell{Average \\ Number \\of Features}  &
\makecell{Average \\Recognition\\Accuracy}  \\
\hline
\hline
CML&\xmark &\xmark &\xmark & \cite{polu2018human,zhang2015recognizing,subasi2018iot,masum2018human,subasi2018sensor} &5&0&94\%\\ \hline
CML&\xmark &\cmark &\xmark & \cite{chen2017robust,ponce2016novel,lago2018improving,siirtola2019user,balli2019human,choi2018temporal,hossain2018improving,elkader2018wearable,voicu2019human,catal2015use,capela2015feature,capela2016evaluation,zubair2016human,yin2015human,weiss2016smartwatch,arif2015physical,capela2015improving,ramos2016combining} &18&19&92\%\\ \hline
CML&\xmark &\cmark &\cmark & \cite{shoaib2016complex,suto2018efficiency,nguyen2019wearable,wang2016comparative,willetts2018semi,de2018human,sukor2018activity,espinilla2018human,su2018activity,rosati2018comparison,mannini2018classifier,davoudi2019accuracy,tian2018adaptive,ding2018energy,attal2015physical,liu2015sensor,pham2015mobirar,zainudin2015activity,zhu2015smartphone,vaizman2017recognizing} &20&56&90\%\\ \hline
CML&\cmark &\xmark &\xmark & \cite{malaise2018activity,zhu2019human} &2&0&94\%\\ \hline
CML&\cmark &\xmark &\cmark & \cite{bhat2018online,lee2016energy} &2&68&88\%\\ \hline
CML&\cmark &\cmark &\xmark & \cite{civitarese2019context,nweke2018analysis,manjarres2018human,ignatov2016human,heng2016human,torres2015robust,de2015multimodal,suarez2015improved,wu2016mixed,lubina2015artificial} &10&13&92\%\\ \hline
CML&\cmark &\cmark &\cmark & \cite{hassan2018robust,chen2018novel,micucci2017unimib,li2019applying,liu2017wearable,badawi2018multimodal,bulbul2018human,cao2018gchar,cruciani2018automatic,san2018robust,mimouna2018human,kwon2018adding,lv2018bi,reyes2016transition,wannenburg2016physical,khalifa2017harke,wang2016triaxial,stisen2015smart,zheng2015human,akhavian2016smartphone,shen2016motion,vavoulas2016mobiact,san2016feature,margarito2015user,davis2016activity,mannini2017activity,damavsevivcius2016human} &27&89&93\%\\ \hline
CML&\cmark &- &- & \cite{vaughn2018activity,paul2015effective,machado2015human,lu2017towards} &4&0&89\%\\ \hline
CML&\xmark &- &- & \cite{rodriguez2017iot,khan2015beyond,liu2015action2activity,abdallah2015adaptive,liu2016action,nguyen2018dealing} &6&0&90\%\\ \hline \hline
DL &\xmark &\xmark &\xmark & \cite{ronao2016human,hammerla2016deep,ravi2016deep,jiang2015human,chen2015deep,murad2017deep,ignatov2018real,alsheikh2016mobile,guan2017ensembles,sathyanarayana2016sleep,ha2016convolutional,wang2018deep,ding2019empirical,nardi2019human,nk2019sensor,inoue2018deep,santos2019accelerometer,milenkoski2018real,zhu2018deep,pienaar2019human,chung2019sensor,zebin2018human,xu2018human,huang2019tse,zhu2019efficient,wang2019perrnn,8727452,zhou2019smartphone,kanjo2019deep,almaslukh2018robust,sun2018sequential,zebin2016human,chen2016lstm,zhang2015human} &34&0&93\%\\ \hline
DL &\xmark &\xmark &\cmark & \cite{zheng2018comparison,lawal2019deep,steven2018feature} & 3&341&92\%\\ \hline
DL &\xmark &\cmark &\xmark & \cite{neverova2016learning,nait2018deep}&2&9&93\%\\ \hline
DL &\xmark &\cmark &\cmark & \cite{uddin2020body,niu2018extreme,wang2016recognition,li2018comparison} & 4&20&92\%\\ \hline
DL &\cmark &\xmark &\xmark & \cite{yang2015deep,ordonez2016deep,grzeszick2017deep,xu2019innohar,jordao2018novel,wang2019attention,panwar2017cnn,alsheikh2016deep} & 8&0&90\%\\ \hline 
DL &\cmark &\cmark &\cmark & \cite{almaslukh2017effective,nutter2018design,zhao2018deep,tao2016multicolumn} & 4&285&90\%\\ \hline

\hline
\end{tabular}}
\end{table*}
\section{Classification Model and Evaluation}\label{sec:eval}
The third and fourth step of the HAR workflow includes identification and evaluation of the classification model that is used for activity recognition. As shown in Figure~\ref{fig:one} and Figure~\ref{fig:three}, CML models still enjoy great popularity compared to those based on the relatively more recent and more advanced models such as the DL models. We point out that many articles made use of different classification models and not just one model for achieving better performance, and as mentioned in Section~\ref{sec:intro} we use accuracy as a comparison metric between the various articles. This beacouse accuracy is the only common metric among them.
\begin{figure}[!thb]
\minipage{0.425\textwidth}
\includegraphics[width=\linewidth,page={11}]{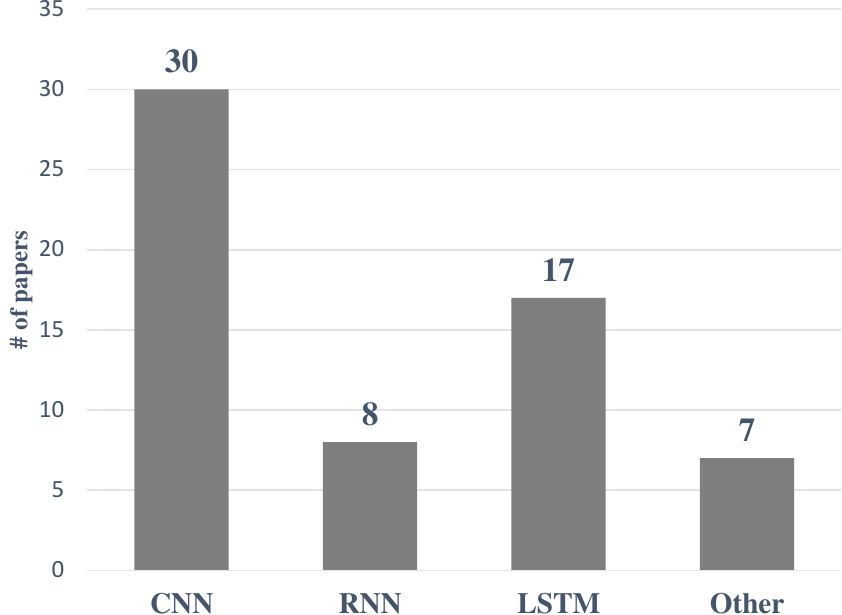} \\ \centering(a)
\endminipage\hfill
\minipage{0.425\textwidth}
\includegraphics[width=\linewidth,page={12}]{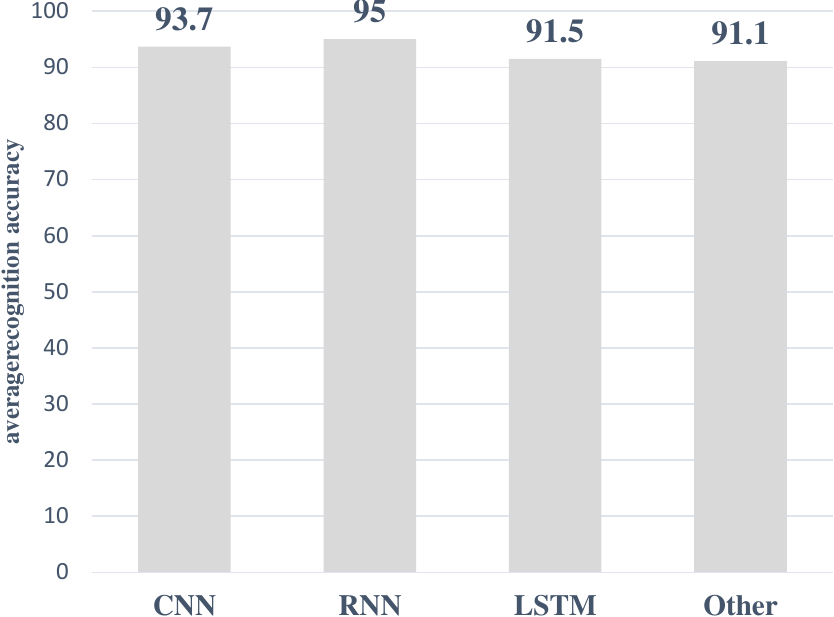} \\ \centering(b)
\endminipage\hfill 
\minipage{0.425\textwidth}
\includegraphics[width=\linewidth,page={13}]{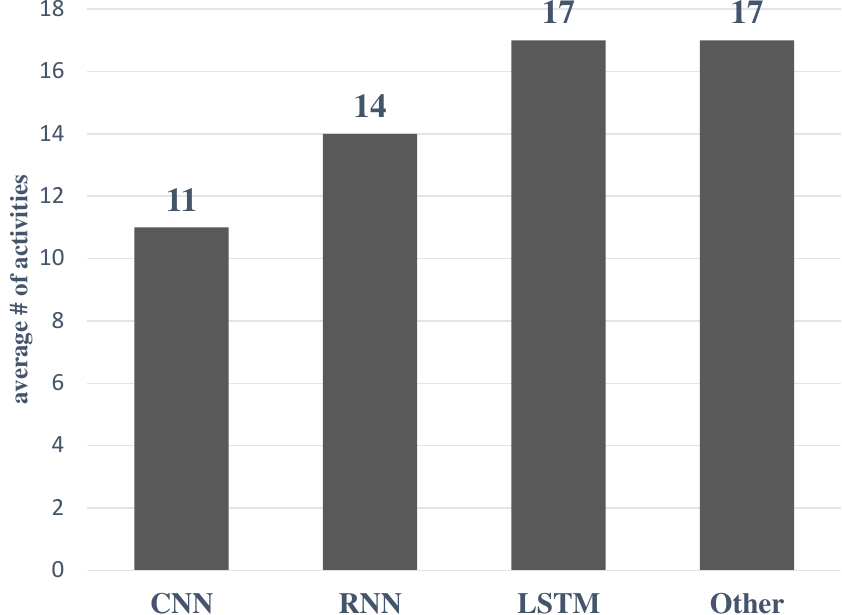} \\ \centering(c)
\endminipage\hfill 
\caption{a) Distribution of Deep Learning Models mostly used in HAR, b) Average activity recognition accuracy of Deep Learning Models mostly used in HAR, and c) Average number of activities of Deep Learning Models mostly used in HAR.}\label{fig:seven}
\end{figure}

\subsection{Deep Learning (DL) based methodologies}
The DL models, as shown in Figure~\ref{fig:one} comprised 54 papers of the 149 papers we reviewed. Figure~\ref{fig:seven} shows (a) the distribution of DL models among the 54 articles,  (b) the average accuracy, and (c) the average number of recognized daily life activities for each model. The most popular model is the Convolutional Neural Network (CNN), which was referenced in 30 papers~\cite{lawal2019deep,nardi2019human,zhu2018deep,huang2019tse,zhu2019efficient,8727452,nutter2018design,wang2019attention,santos2019accelerometer,xu2018human,niu2018extreme,ding2019empirical,zhou2019smartphone,sun2018sequential,almaslukh2018robust,zheng2018comparison,kanjo2019deep,jordao2018novel,grzeszick2017deep,panwar2017cnn,li2018comparison,zebin2016human,sathyanarayana2016sleep,ha2016convolutional,ravi2016deep,ignatov2018real,chen2015deep,jiang2015human,ronao2016human,yang2015deep}. The CNN models obtained an average accuracy of 93.7\% in activity recognition over an average number of 11 activities of daily life. The second most used model was the Long Short-Term Memory (LSTM) model, which was used in 17 papers~\cite{sun2018sequential,kanjo2019deep,pienaar2019human,steven2018feature,murad2017deep,li2018comparison,sathyanarayana2016sleep,zebin2018human,chung2019sensor,milenkoski2018real,nait2018deep,zhao2018deep,chen2016lstm,tao2016multicolumn,guan2017ensembles,hammerla2016deep,ordonez2016deep}. It obtained an average accuracy of 91.5\% over an average number of 17 activities of daily life. 
Recurrent Neural Network (RNN) were used in~\cite{sathyanarayana2016sleep,hammerla2016deep,uddin2020body,inoue2018deep,neverova2016learning,murad2017deep,wang2019perrnn,pienaar2019human}, over an average number of 14  obtaining an average accuracy of 95\%. 
Finally, the rest of the papers (indicated by Other in Figure~\ref{fig:seven}) where based on models such as Autoencoders~\cite{almaslukh2017effective,wang2016recognition}, Inception Neural Networks (INN), or the other frameworks~\cite{alsheikh2016mobile} for a total of 7 papers with an average accuracy of 91.1\% and an average number of 17 activities of daily life.

\begin{figure}[!t]
\minipage{0.45\textwidth}
\includegraphics[width=\linewidth,page={14}]{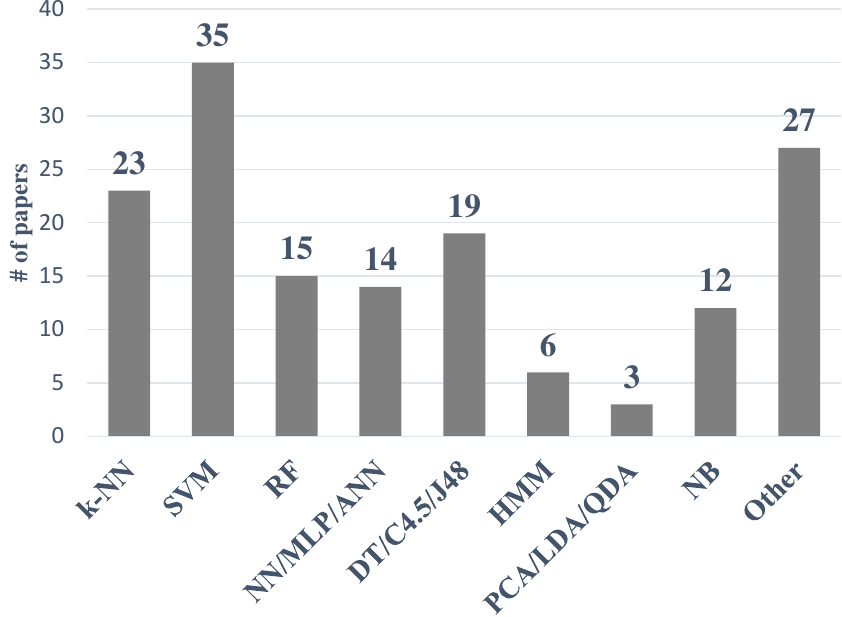} \\ \centering(a)
\endminipage\hfill
\minipage{0.45\textwidth}
\includegraphics[width=\linewidth,page={15}]{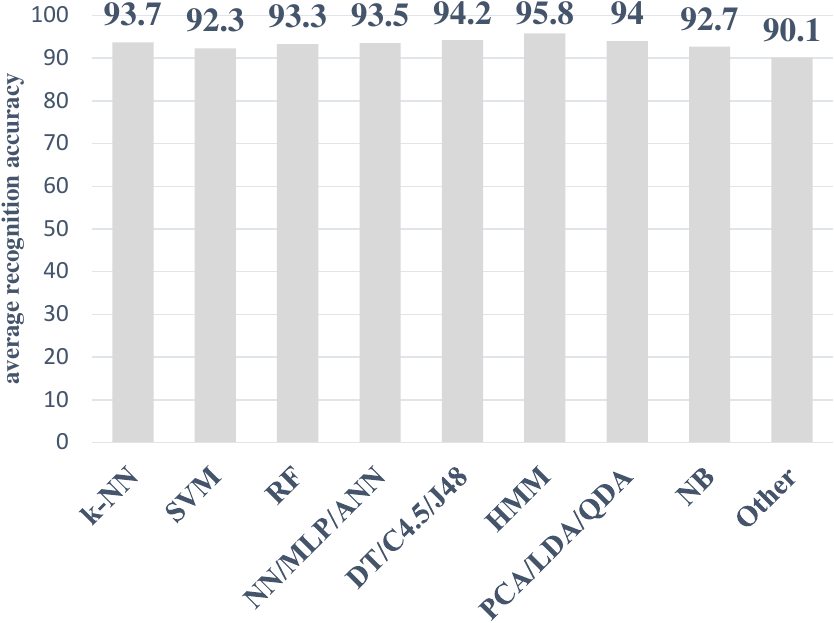} \\ \centering(b)
\endminipage\hfill 
\minipage{0.45\textwidth}
\includegraphics[width=\linewidth,page={16}]{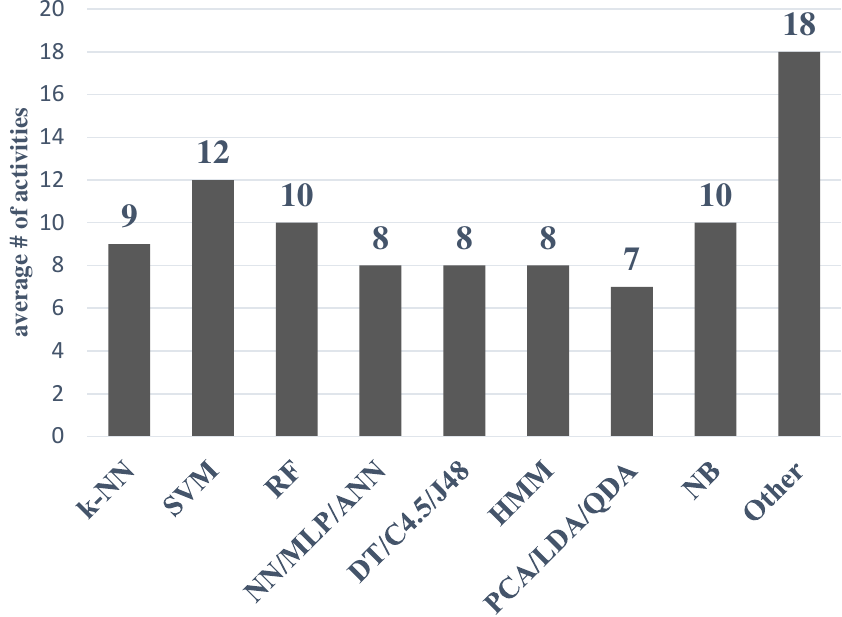} \\ \centering(c)
\endminipage\hfill 
\caption{a) Distribution of CML Models mostly used in HAR, b) Average activity recognition accuracy of CML Models mostly used in HAR, and c) Average number of activities of CML Models mostly used in HAR.}\label{fig:eight}
\end{figure}

\subsection{Machine Learning (ML) based methodologies}
Among the 149 reviewed papers, as shown in Figure~\ref{fig:one}, 95 presented an HAR methodology based on classical ML. Figure~\ref{fig:eight} shows (a) the distribution of these models, (b) the obtained average accuracy and (c) the average number of recognized activities of daily life. Among the different types of classical ML models, the most commonly used model was the Support Vector Machine (SVM) model ~\cite{masum2018human,bulbul2018human,zhu2019human,rosati2018comparison,espinilla2018human,mimouna2018human,balli2019human,tian2018adaptive,de2018human,subasi2018iot,cruciani2018automatic,zhu2015smartphone,liu2017wearable,heng2016human,lee2016energy,torres2015robust,yin2015human,zheng2015human,mannini2017activity,davis2016activity,damavsevivcius2016human,akhavian2016smartphone,stisen2015smart,reyes2016transition,attal2015physical,liu2015action2activity,liu2016action,lago2018improving,choi2018temporal,davoudi2019accuracy,elkader2018wearable,mannini2018classifier,ramos2016combining,chen2017robust,micucci2017unimib} which was used in 35 papers, achieving an average accuracy of 92.3\% over an average of 12 activities. 
The second most used model is the classical k-Nearest Neighbor (kNN) model~\cite{masum2018human,bulbul2018human,rosati2018comparison,espinilla2018human,balli2019human,de2018human,subasi2018iot,cruciani2018automatic,zhu2015smartphone,liu2017wearable,torres2015robust,akhavian2016smartphone,stisen2015smart,attal2015physical,liu2015action2activity,liu2016action,manjarres2018human,paul2015effective,ignatov2016human,arif2015physical,wang2016comparative,shoaib2016complex,vaughn2018activity}, which was used in 23 papers, achieving an average accuracy of 93.7\% over an average of 12 activities of daily life. 
The third and fourth most used model are the Decision Tree (DT) model~\cite{rosati2018comparison,kheirkhahan2019smartwatch,espinilla2018human,balli2019human,de2018human,subasi2018iot,yin2015human,akhavian2016smartphone,stisen2015smart,shoaib2016complex,pham2015mobirar,capela2016evaluation,capela2015feature,rodriguez2017iot,zainudin2015activity,vavoulas2016mobiact,weiss2016smartwatch,davoudi2019accuracy,tian2018adaptive,ramos2016combining}, which was used in 19 papers, obtaining an average accuracy of 94.2\% over an average of 8 activities of daily life, and the Random Forest (RF)~\cite{masum2018human,balli2019human,subasi2018iot,cruciani2018automatic,stisen2015smart,attal2015physical,manjarres2018human,weiss2016smartwatch,ding2018energy,badawi2018multimodal,nweke2018analysis,polu2018human,shen2016motion,micucci2017unimib,davoudi2019accuracy}, which was used in 15 papers, obtaining an average accuracy of 93.3\% over an average of 10 activities of daily life.
The fifth most used model is the Neural Networks (NN) ~\cite{rosati2018comparison,lubina2015artificial,suto2018efficiency,espinilla2018human,yin2015human,catal2015use,de2018human,subasi2018iot,cruciani2018automatic,akhavian2016smartphone,liu2016action,weiss2016smartwatch,nguyen2018dealing,voicu2019human}, which was used in 14 papers, obtaining an average accuracy of 93.5\% over an average of 8 activities of daily life. 
Other used models are the Naïve Bayes (NB)~\cite{vaughn2018activity,wang2016comparative,rodriguez2017iot,capela2015feature,weiss2016smartwatch,espinilla2018human,de2018human,torres2015robust,yin2015human,zheng2015human,liu2016action,ramos2016combining}, the Dynamic Bayesian Network (DBN)~\cite{zhang2015recognizing,hassan2018robust,alsheikh2016deep},
\ac{HMM}~\cite{davis2016activity,attal2015physical,pham2015mobirar,malaise2018activity,san2018robust,san2016feature}, Extreme Learning Machine (ELM)~\cite{niu2018extreme,sun2018sequential}, \ac{PCA}, \ac{LDA}, \ac{QDA}~\cite{sukor2018activity,siirtola2019user,chen2017robust} and many others~\cite{civitarese2019context,nguyen2019wearable,li2019applying,ponce2016novel,subasi2018sensor,kwon2018adding,willetts2018semi,su2018activity,hossain2018improving,chen2018novel,lv2018bi,capela2015improving,cao2018gchar,suarez2015improved,khan2015beyond,khalifa2017harke,wang2016triaxial,wu2016mixed,machado2015human,liu2015sensor,micucci2017unimib,wannenburg2016physical,margarito2015user,abdallah2015adaptive,lu2017towards}. 
It is noteworthy that some of the articles have tested their approaches using different models.
\section{Discussion}\label{sec:disc}
In this paper, we provided an overview of the current HAR research. HAR is a critical research area in activity recognition, pervasive computing, and human assistive environments. In the last decades, with the rise of new technologies and with growing needs such as aging population, HAR is becoming even more essential. In recent years, DL-based HAR methods have produced excellent results in terms of recognition performance. However, CML-based approaches are still widely used, and they generate outstanding results without the computational costs.
\begin{figure}[!bh]
\centering
\includegraphics[width=0.45\textwidth,page={17}]{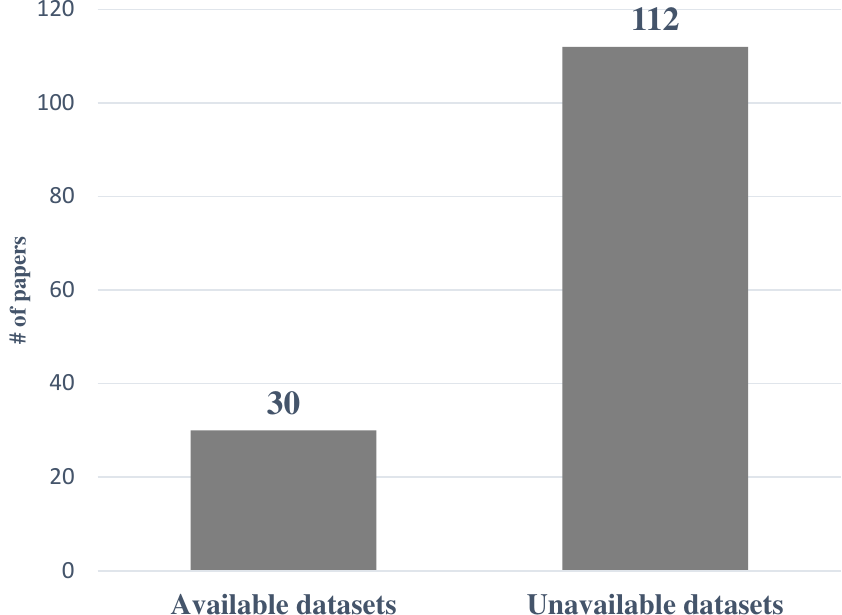}
\caption{Availability of datasets used to evaluate the proposed methodologies.}\label{fig:nine}
\end{figure}
However, in recent years, the reproducibility of ML models has become increasingly important. Based on our research, for 78\% of the proposed HAR methodologies, the results are not fully reproducible due to proprietary datasets. This results in barriers for the research community for the identification of the best models and benchmarking the results.
As shown in Figure~\ref{fig:nine}, starting from the initial 293 papers and after the removal of surveys and on payment articles, among a total of 142 datasets, only 30 datasets are publicly available, some of which are shown in the Table~\ref{tab:five}.
\begin{figure}[!thb]
\centering
\includegraphics[width=0.45\textwidth,page={18}]{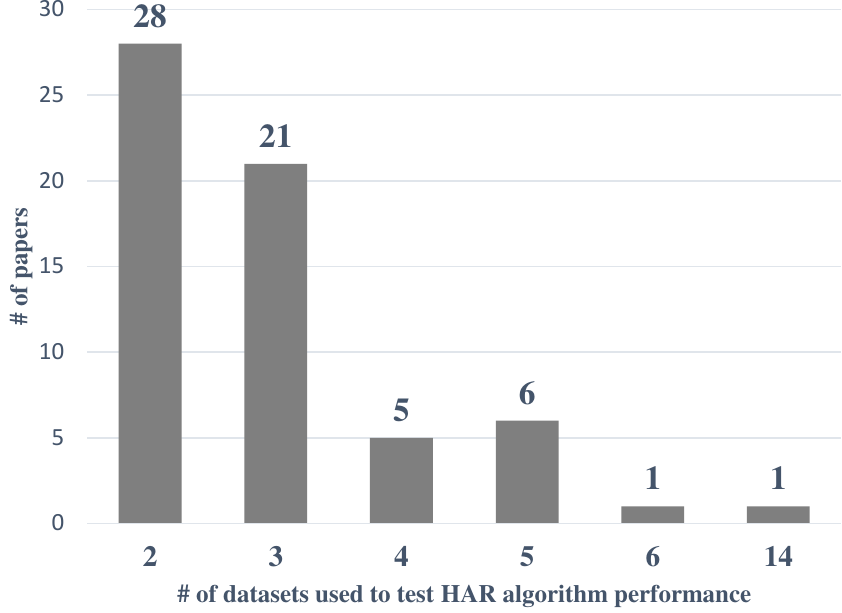}
\caption{Number of datasets (x-axis) used to test an article and number articles methodologies (y-axis) tested on such number of datasets (166 articles were tested on only one dataset).}\label{fig:ten}
\end{figure}

\begin{figure*}[!htb]
\centering
\includegraphics[width=0.8\textwidth,page={19}]{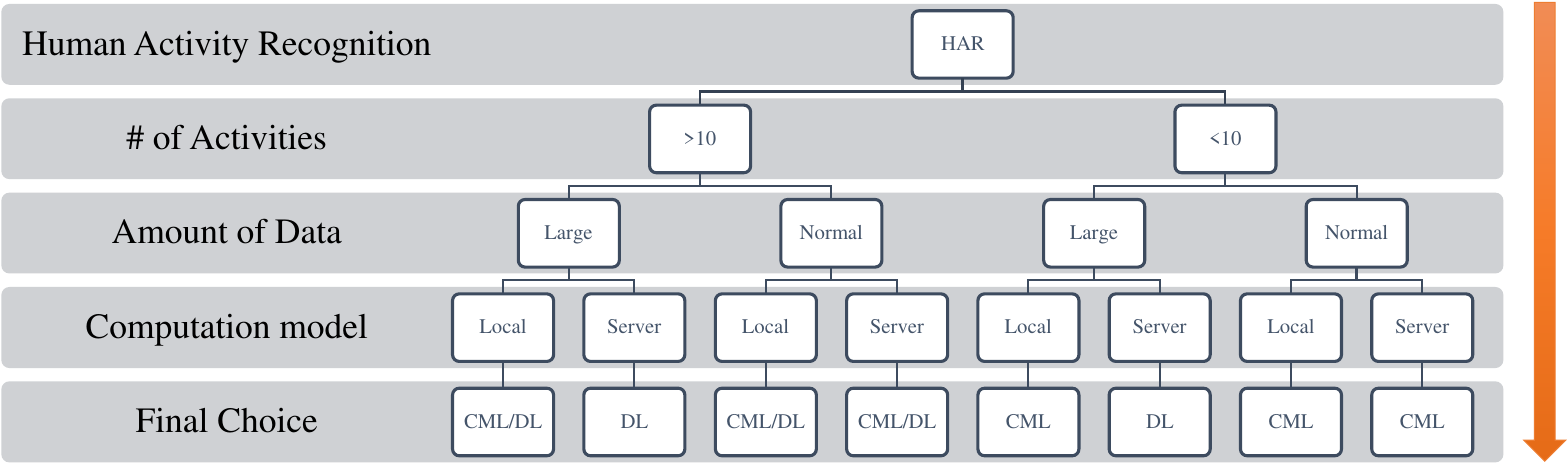}
\caption{Model selection diagram. DL=Deep Learning, CML= Classic Machine Learning }\label{fig:eleven}
\end{figure*}
Furthermore, the lack of public heterogeneous datasets reduces the possibility of creating HAR models with better generalization capabilities.
This is because the data used in the investigated papers are collected primarily in a controlled environment. This problem is exacerbated by the inter-subject and intra-subject variability absent in such scripted datasets, as most proposed HAR models are only tested on a limited number of activities and captured in a single controlled environment. 
Among the 149 analyzed HAR models, 87 models were tested on a single dataset, with the remaining 62 tested on more than one dataset. 
As shown in Figure~\ref{fig:ten}, we found that 28 HAR methodologies were tested on two datasets, 21 HAR methodologies on three datasets, less than 10 HAR methodologies on 4-6 datasets, and only one methodology~\cite{janidarmian2017comprehensive} was tested on a total of 14 datasets. This situation shows the challenge of identifying a methodology superior to the others. 

Another significant issue concerns the interpretability of the results, mainly related to papers presenting similar methodologies and tested on the same dataset, claiming to achieve almost the same results in terms of activity recognition accuracy. Such an issue is related to tests performed using commercial tools, lack of open source code, and authors who do not publicly provide their source code. 
Besides, the heterogeneity of the data and the definition of a HAR methodology that can recognize the activities carried out by people with different physical and motor characteristics collides with the data sources used for data collection. As we have seen, a variety of sensors and devices are used for data collection. However, the proposed methodologies are usually very rigid regarding the data source.
Specifically, it becomes difficult to have a methodology tested on a particular individual by making use of a particular sensor(s) and subsequently changing the sensor model. Various sensors have different technical characteristics, which also entail their specific state, e.g., the measurement error or the noise that a specific sensor presents.

Regarding the HAR models, Figure~\ref{fig:seven} and Figure~\ref{fig:eight} show that CML models are still used more widely than complex DL-based models. This is because CML models require a smaller amount of training data, as well as lower computational requirements. In addition, DL models are inherently difficult to interpret. Nonetheless, DL models have a unique ability to recognize more complex activities, while maintaining high accuracy. In addition, they do not require a data preprocessing stage. 
Figure~\ref{fig:eleven} shows a suggested workflow for developing HAR applications based on:
\begin{itemize}
    \item the number of activities to be recognized,
    \item the amount of available (labeled) data,
    \item local or remote computation.
\end{itemize}
We observed that the selection of the precise DL or CML model is primarily based on the computational requirements and the amount of available training (labeled) data. In terms of the sensors, the most widely used used, if not indispensable, sensor is the accelerometer, which can be used in conjunction with other sensors such as the gyroscope or the magnetometer.
\section{Future Research Direction}\label{sec:future}
Based on reviewed papers, a few possible research directions are noted below. One of the main limitations of HAR algorithms is the lack of standardized methodologies that can generalize to heterogeneous set of activities performed by a diverse set of users. As a potential solution, transfer learning could reuse the knowledge acquired in one problem to solve a similar problem. For example, knowledge acquired based on a specific inertial sensor positioned on a specific body location can potentially be reused with a different sensor location or with a different type of inertial sensor. The extent to which transfer learning can be helpful in various scenarios, is not investigated in a comprehensive manner and needs to be further studied. Sensor fusion also provides a promising path. In particular, merging different sensors could address issued related to reliability and accuracy of a single sensor and could also enrich collected information. When data from one modality is not reliable, the system could switch to a different sensor modality to ensure robust data collection. Another research direction is fine-grained activity recognition based on examining daily object interactions. This will allow us to recognize sub-actions and sequence of actions and will provide much richer context information to downstream applications. Sensor fusion can also be helpful when a large number of inertial sensors or proximity sensors are attached to daily objects. 
To further advance the progress in this area, we provide a set of recommendations. First, developing benchmark datasets should be a priority for the HAR community. New HAR models should be compared with available HAR models on benchmark data to show improvement. Furthermore, creation of datasets with an adequate number of subjects and diverse set of activities is strongly recommended. Fine-grained activity recognition also could benefit from large-scale, standardized benchmarks. Researchers working on HAR algorithms should also pay attention to hardware and system issues, besides solely developing and improving HAR algorithms. On-device computation should be a primary goal, as well as analysis of memory, CPU, and battery consumption, to explore the trade-off between resource utilization and recognition accuracy. Finally, position and orientation dependence should be extensively studied; otherwise, the design of position/orientation-dependent techniques could result in inconsistent and non-robust downstream applications.
\section{Conclusion}\label{sec:conc}
\begin{figure}[!thb]
\centering
\includegraphics[height=0.25\textheight,width=0.45\textwidth,page={22}]{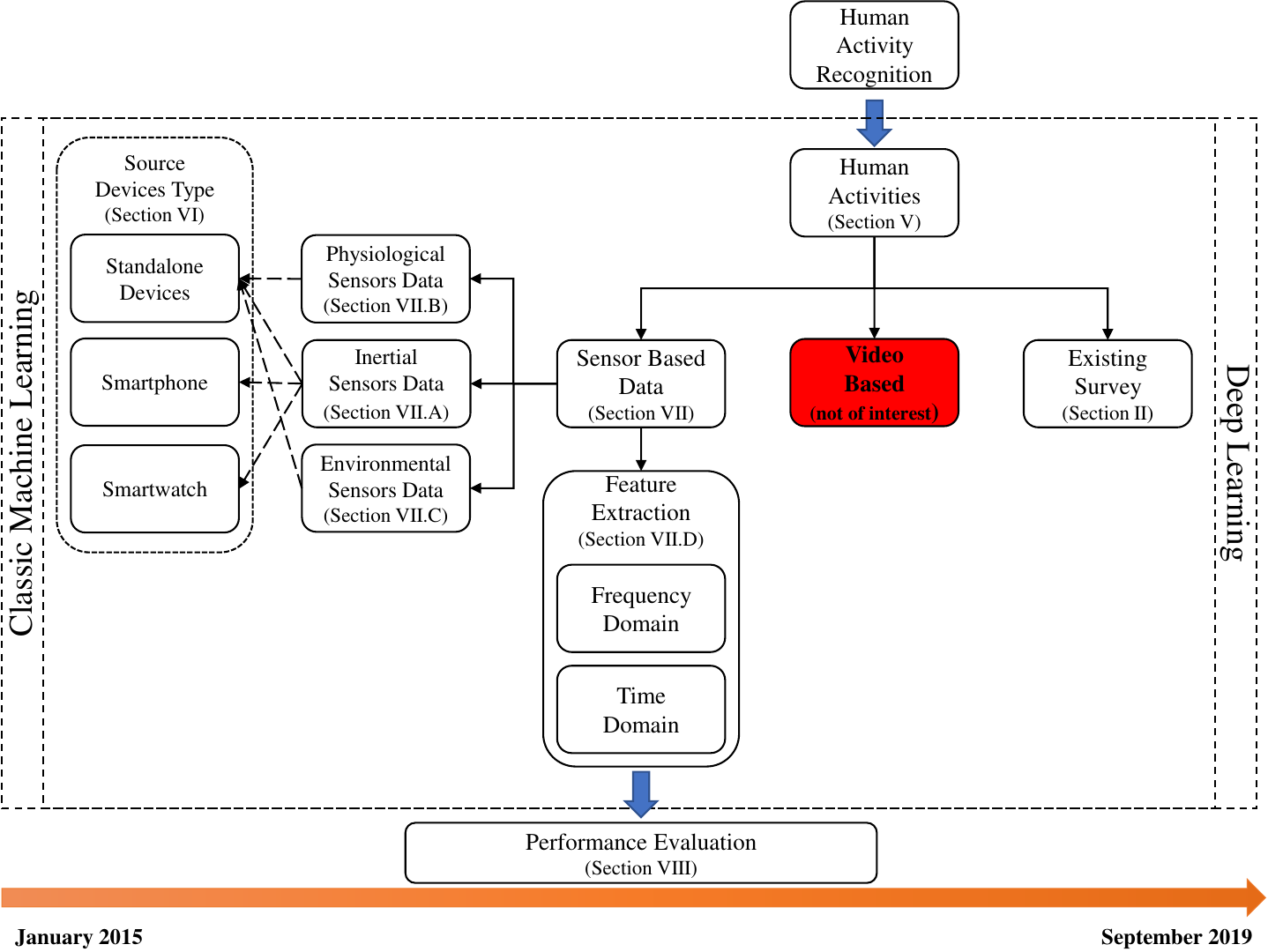}\caption{Overview of the proposed survey structure on sensor-based HAR research results from 2015 to 2019.}\label{fig:twelve}
\end{figure}

HAR systems have become a growing research area in the past decade, achieving impressive progress. In particular, sensor-based HAR have many advantages compared to vision-based HAR methodologies, which pose privacy concerns and are constrained by computational requirements. Activity recognition algorithms based on ML and DL are becoming central in HAR. Figure~\ref{fig:twelve} summarizes HAR methodologies between January 2015 and September 2019.
Starting from a meta-review of the existing HAR surveys, we analyzed the reviewed literature based on the most widely studied human activities (Section V), the most used electronic sensors as the data source (Section VII), and the most known devices that integrate with these sensors (Section VI) without taking into account the video-based methodologies. In detail, sensor-based data perceived by physiological, inertial, and environmental sensors were of primary interest. Device types were also extensively studied, categorizing them in: a) standalone, b) smartphone, and c) smartwatch devices. For each category, results were shown in terms of the average number of recognized activities, the average number of datasets used to test the methodologies, and the average accuracy. This survey also dis-cussed methodologies based on accelerometer, gyroscope, and magnetometer. We also discussed the preprocessing approaches and their results based on feature extraction, noise removal, and normalization techniques.
Moreover, we discussed datasets primarily in the literature, emphasizing publicly available datasets.
Finally, we presented a description of the recognition models most used in HAR. 
For this purpose, we have presented the most widely used DL and ML models and their results, both from the point of view of quality (accuracy) and quantity (number of recognized activities). 
\begin{figure*}[!thb]
\centering
\includegraphics[height=0.275\textheight,width=1\textwidth,page={23}]{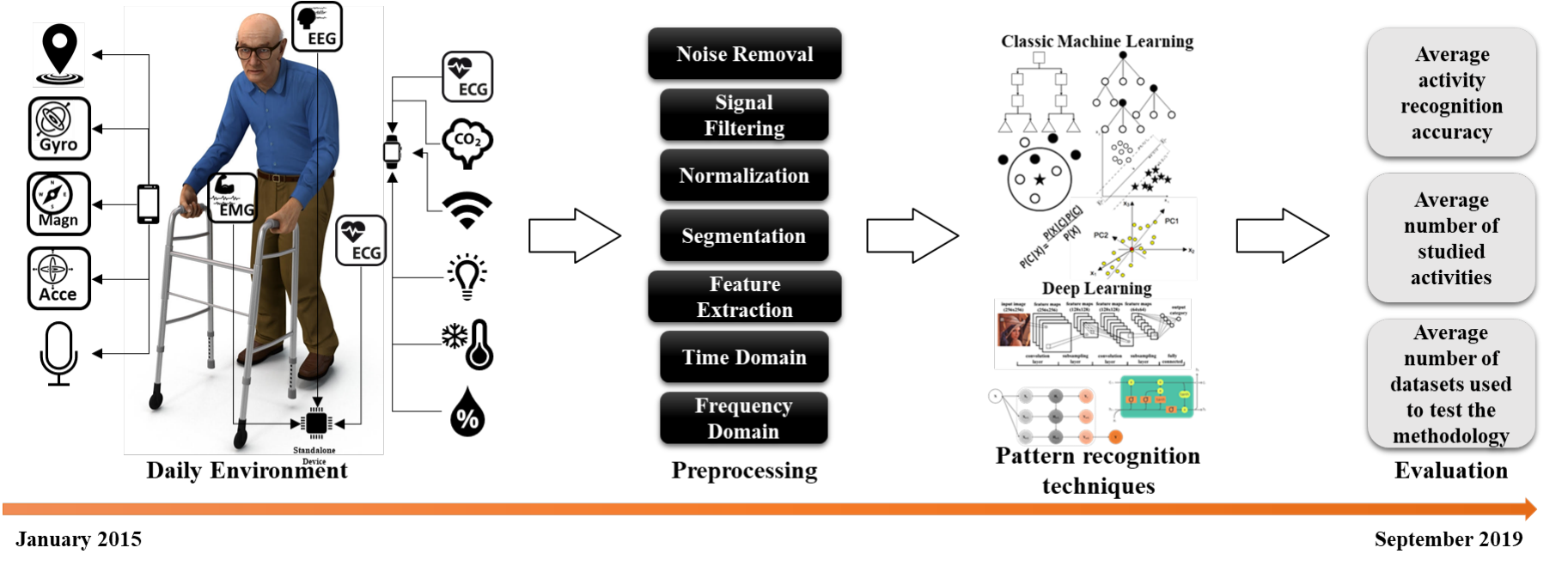}\caption{A Systematic Review of Human Activity Recognition (HAR) approaches, published from January 2015 to September 2019, based on Classical Machine Learning (CML) and Deep Learning (DL), which make use of data collected by sensors (inertial or physiological), embedded into wearables or environment. We surveyed methodologies based on sensor type, device type (smartphone, smartwatch, standalone), preprocessing step (noise removal/feature extraction technique), and finally, their DL or CML model. The results are presented in terms of a) average activity recognition accuracy, b) the average number of studied activities, and c) the average number of datasets used to test the methodology.}\label{fig:fifty}
\end{figure*}
We concluded that HAR researchers still prefer classic ML models, mainly because they require a smaller amount of data and less computational power than DL models. However, the DL models have shown higher capacity in recognizing many complex activities. Future work should focus on the development of methodologies with more advanced generalization capabilities and recognition of more complex activities.
To summarize, Figure~\ref{fig:fifty} shows a Graphical Abstract (GA) of the workflow of this survey.
\section{ACKNOWLEDGMENT}
A.B. and P.R. were supported by R01 GM110240 from the National Institute of General Medical Sciences. P.R. has received grant NIH/NIBIB 1R21EB027344 and NSF CAREER 1750192.  The content is solely the responsibility of the authors and does not necessarily represent the official views of the National Institutes of Health or National Science Foundation.

\ifCLASSOPTIONcaptionsoff
  \newpage
\fi

\linespread{0.9}
\bibliographystyle{IEEEtran}
\bibliography{IEEEabrv,biblio}{}
\begin{IEEEbiography}
[{\includegraphics[width=1in,height=1.25in,clip,keepaspectratio]{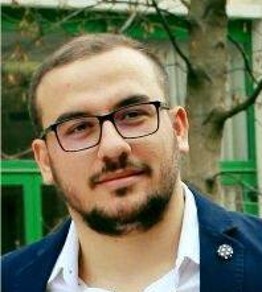}}]{\textbf{Florenc Demrozi,}} PhD in computer science, IEEE member, received the B.S. and M.E. degrees in Computer Science and Engineering from the University of Verona, Italy, respectively in 2014 and 2016, and the Ph.D. degree in Computer Science from University of Verona, Italy, in 2020. He is currently a Postdoctoral researcher and Temporary Professor at the Department of Computer Science, University of Verona, Italy, where he is member of the ESD (Electronic Systems Design) Research Group, working on Ambient Intelligence (AmI), Ambient Assisted Living (AAL) and Internet of Things (IoT).
\end{IEEEbiography}
\begin{IEEEbiography}
[{\includegraphics[width=1in,height=1.25in,clip,keepaspectratio]{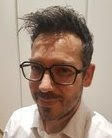}}]{\textbf{Graziano Pravadelli,}} PhD in computer science, IEEE senior member, IFIP 10.5 WG member, is full professor of information processing systems at the Computer Science Department of the University of Verona (Italy) since 2018. In 2007 he cofounded EDALab s.r.l., an SME working on the design of IoT-based monitoring systems. His main interests focus on system-level modeling, simulation and semi-formal verification of embedded systems, as well as on their application to develop IoT-based virtual coaching platforms for people with special needs. In the previous contexts, he collaborated in several national and European projects and he published more than 120 papers in international conferences
and journals.
\end{IEEEbiography}
\begin{IEEEbiography}
[{\includegraphics[width=1in,height=1.25in,clip,keepaspectratio]{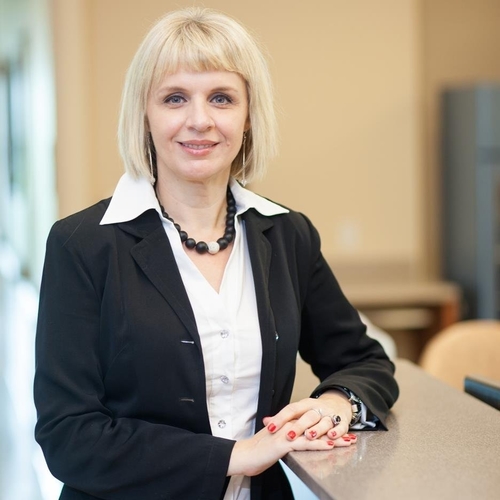}}]{\textbf{Azra Bihorac,}} MD MS FASN FCCM is a R. Glenn Davis Professor of Medicine, Surgery, Anesthesiology and Physiology and Functional Genomics at University of Florida. She leads Precision and Intelligence in Medicine Partnership (PrismaP), a multidisciplinary research group of experts in data science and informatics, focused on the development and implementation of intelligent systems and technologies to augment clinical decisions and optimize health care delivery in surgery, critical care medicine and nephrology. The team is developing machine learning and informatics tool for real-time risk stratification and annotation of hospital-acquired complications and kidney disease as well as for the application of omics technologies on urine for predictive enrichment of patients with critical illness. Her vision is to develop tools for intelligent human-centered health care that delivers optimized care tailored to a patient’s “personal clinical profile” using digital data. Through her work in national and international professional organizations in nephrology and critical care medicine, she has advocated for women physicians and scientists, promoting their equality and recognition in health care leadership, research and education. She completed her MD at the University of Sarajevo, Bosnia and Herzegovina, internal medicine residency at Marmara University, Istanbul, Turkey and University of Florida, fellowships in critical care medicine and nephrology and her Masters in Clinical Science at University of Florida.
\end{IEEEbiography}
\begin{IEEEbiography}
[{\includegraphics[width=1in,height=1.25in,clip,keepaspectratio]{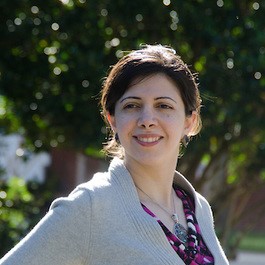}}]{\textbf{Parisa Rashidi}}
received her PhD in computer science in 2011 with an emphasis on machine learning. She is currently an associate professor at the J. Crayton Pruitt Family Department of Biomedical Engineering (BME) at University of Florida (UF). She is also affiliated with the Electrical \& Computer Engineering (ECE), as well as Computer \& Information Science \& Engineering (CISE) departments. She is the director of the “Intelligent Health Lab” (i-Heal).  Her research aims to bridge the gap between machine learning and patient care. She has served on the technical program committee of several conferences and has been a reviewer of numerous IEEE journals.
\end{IEEEbiography}
\end{document}